\documentclass[acmsmall,nonacm]{acmart}
\usepackage{multirow}
\usepackage{todonotes}
\usepackage{colortbl}
\usepackage{xcolor}
\newcommand{\added}[1]{\textcolor{black}{#1}}
\newcommand{\final}[1]{\textcolor{black}{#1}}

\AtBeginDocument{%
  }

\setcopyright{acmcopyright}
\copyrightyear{2023}
\acmYear{2023}
\acmDOI{10.1145/36351}

\setcopyright{acmlicensed}
\acmJournal{TOCHI}
\acmYear{2023} \acmVolume{1} \acmNumber{1} \acmArticle{1} \acmMonth{1} \acmPrice{15.00}\acmDOI{10.1145/3635147}

\author{Hendrik Heuer}
\email{hheuer@uni-bremen.de}
\orcid{0000-0003-1919-9016}
\affiliation{%
 \institution{Harvard University}
 \city{Cambridge, MA}
 \country{USA}
 }
\affiliation{%
 \institution{University of Bremen}
 \city{Bremen}
 \country{Germany}
 }

\author{Elena L. Glassman}
\email{glassman@seas.harvard.edu}
\orcid{0000-0001-5178-3496}
\affiliation{%
 \institution{Harvard University}
 \city{Cambridge, MA}
 \country{USA}
 }

\begin{document}

\title{Reliability Criteria for News Websites}

\begin{abstract}
Misinformation poses a threat to democracy and \final{to} people's health. Reliability criteria for news websites can help people identify misinformation. \final{But d}espite their importance, \final{there has been no} empirically substantiated list of criteria for \added{distinguishing} reliable from unreliable news websites. We identify reliability criteria, describe how they are applied in practice, and compare them to prior work. Based on \final{our} analysis, we distinguish \added{between manipulable and less manipulable criteria} and compare politically diverse laypeople \added{as end users} and journalists \added{as expert users}. We discuss \final{11} \added{widely recognized} criteria, including the following \final{6} criteria that are \final{difficult} to manipulate: content, political alignment, authors, professional standards, what sources are used, and \final{a website's reputation}. Finally, we describe how technology may be able to support people in applying \final{these} criteria in practice to assess the reliability of websites.
\end{abstract}

\ccsdesc[500]{Human-centered computing~Empirical studies in collaborative and social computing}
\ccsdesc[500]{Human-centered computing~Empirical studies in HCI}

\keywords{Misinformation, Disinformation, Fake News, Criteria, Media Litearcy, Checklist, Online News, Fact-checking, Credibility, Information Quality}

 \maketitle

\section{Introduction}

Many people find it challenging to distinguish reliable from unreliable information. \final{Misinformation played an important role in the COVID-19 pandemic~\cite{world_health_organization_managing_2020}, the 2021 United States Capitol attack~\cite{painter2022big}, and the Russian invasion of Ukraine~\cite{thompson_alba_2022}.} In their large-scale study, Arechar et al. showed that misinformation is a global problem, and that the psychological factors that underlie misinformation are similar across the globe~\cite{rand16countries}. In the U.S., nine out of ten adults believe that false information is a challenging problem~\cite{Mitchell_Pew_2019}, and a Pew Research Center study showed that most people in the U.S. have difficulty distinguishing real news from opinions~\cite{mitchell2018can}.

Reliability criteria \final{can} empower readers to be more discerning when \final{navigating} a landscape that may include misinformation. \added{The goal of this paper was to yield robust news reliability criteria independent of a person's experience with how news is produced. This paper is the first to empirically ground such criteria in a study that combines elements of \final{contextual inquiry~\cite{salazar2020contextual}, \textit{think aloud} study~\cite{kuusela2000comparison}, and semi-structured interviews~\cite{kuniavsky2003observing}. Our study focuses on reliability criteria for end users of genuine, live news websites with many articles}. We report all criteria used by participants while evaluating different news websites in two countries, and discuss why certain criteria are more difficult than others to manipulate.}

\final{It is important to investigate news websites in a realistic \final{and live} setting, since four} out of five people~(82\%) worldwide use online sources to read news~\cite{newman_reuters_2022}. Almost three out of five~(57\%) rely on social media, where links to online websites are frequently shared. For both online websites in general and social media in particular, there is little guidance on distinguishing reliable from unreliable news websites. \final{While prior research showed that source labels are effective~\cite{10.1145/3415211,10.1145/3491102.3517717}, labels for entire websites are rarely provided. The reliability criteria for news websites presented in this paper can help users recognize websites that, while appearing to provide news, may not adhere to journalistic standards and may provide biased or false information for political or monetary gain.}

\final{In the context of misinformation, Wardle distinguishes between seven types of mis- and disinformation~\cite{wardle_2018}: 1)~misleading content, 2)~false connection, 3)~false context, 4)~manipulated content, 5)~fabricated content that is 100\% false, 6)~imposter content impersonating genuine sources, and 7)~satire. While applicable to all seven types, the reliability criteria for news websites presented in this paper are especially relevant for online websites that provide misleading content, false connection, false context, and manipulated content.}

\final{A person's partisanship and political stance have been shown to influence their assessment of news reliability. In the context of the crowdsourced fact-checking program on \final{X, formerly Twitter,} for example, Allen et al. found that users tend to selectively question or dispute content posted by people with whom they disagree politically~\cite{10.1145/3491102.3502040}. Work by Pereira et al. indicates that political identity increases the likelihood of a person believing positive news about their ingroup and negative news about their outgroup~\cite{pereira2023identity}. Pennycook et al., on the other hand, found that people's assessment of the plausibility of headlines is independent of whether a story is consistent or inconsistent with their political ideology~\cite{pennycook2019lazy}. Overall, prior work indicates that partisanship plays a complex role in the context of misinformation, motivating us to account for partisanship in our study design. A unique aspect of our work is covering the political spectrum in two countries. We provide a comprehensive overview of the criteria used in practice by professional journalists as expert users and laypeople \final{end users} across political affiliations.} \added{The term \textit{end user}, in this context, refers to people without specialized knowledge about how to assess whether a news website is a reliable source to inform themselves and others.} 

\final{While} initial steps towards reliability criteria have been made\added{~\cite{doi:10.1073/pnas.1920498117,newsguard_2022,10.1145/3415164,zhang2018structured,bradshaw2020sourcing}}, and \final{while} research indicates that these criteria are helpful in practice\added{~\cite{rand16countries}}, which criteria users \added{can} apply and \final{which are useful in practice remain open questions}. Based on observing politically diverse laypeople \added{as end users} and journalists \added{as expert users}, \final{we provide news reliability criteria that focus not on individual articles, but on entire news websites.} 

\final{With far fewer websites than articles, providing reliability criteria for entire websites saves time, as users do not have to evaluate individual news articles and claims. The cost of starting a news website is also higher than the cost of publishing and promoting individual articles. These criteria also mean it is not necessary to engage with individual claims in a story to explain why a website is unreliable, and they can support users before an article containing misinformation is even created and be integrated more easily into social media platforms.}

To observe which reliability criteria are actually applied in practice, \added{we studied professional journalists as experts and politically diverse laypeople as end users. \final{With knowledge of  how news is produced, professional journalists are experts in assessing the reliability of news websites and in fact-checking claims. End users have less experience with how news is produced. By looking at both experts and end users, we identify criteria 1)~that are used by \added{experts}, 2)~that are within \final{end users}' abilities, and 3)~that debunk common \final{end users'} choices that are less helpful for identifying reliable information. As explained, prior work has shown} that the assessment of news reliability can be influenced by a person's partisanship and political stance~\cite{10.1145/3491102.3502040,pennycook2019lazy, pereira2023identity}.} To do justice to this complexity, we recruited elected representatives from a state parliament in Germany as \final{end users}. The political alignment of elected representatives can be clearly identified and, as a group, they are representative of the many different common political stances within \final{the society}. Our sample represents the voting body in the Free Hanseatic City of Bremen, one of 16 federal states in Germany.

\added{\final{We used} eleven criteria to determine the reliability of online information: content, political alignment, writing style, authors, self-description, professional standards, advertisements, ownership, sources, reputation, and \final{website design}. These criteria} are intended to directly help news consumers assess the reliability of news websites they encounter on social media and through search engines. The criteria may also be helpful for those who operate social media platforms like TikTok, Facebook, and Telegram\final{, who could use the criteria to streamline content moderation processes, train content moderators, and show our criteria to users}. \added{In addition, the criteria may support those who witness the sharing of misinformation, e.g., by helping them spot the misinformation and \final{justify why it qualifies as such.}} Witnesses could include journalists who fact-check a particular website or people who encounter friends or family members sharing questionable links, e.g., in a family group chat. For these situations, we hope \final{our} reliability criteria for news websites' \added{practice} are a quick and empirically sound way to decide and explain why a given website may be unreliable.

We distinguish between \added{criteria that can be easily manipulated by misinformation providers and other malicious actors and criteria that are more difficult to manipulate}. \added{In Section~\ref{sec:discussion_comparison_prior_work}, we also compare the criteria to prior work based on different methodologies. \final{We conclude by explaining how the criteria we consider difficult to manipulate}} can be best implemented in practice.

We address the described gaps in the literature by making the following contributions:

\begin{itemize}
    \item We examine how \added{end users and experts} assess the reliability of news websites through \added{an empirical investigation that combines elements of contextual inquiry, think aloud study, and semi-structured interviews}~(RQ1).
    \item We \added{present} criteria based on \added{our empirical work} that could be used by future news consumers to help them assess the reliability of news websites~(RQ2).
    \item We compare the similarities and differences \final{of end users and experts} with respect to identifying misinformation~(RQ3).
\end{itemize}

\section{Background}

With the advent \final{and expansion} of social media, the distribution of misinformation has become faster and cheaper. Consequently, misinformation has received \final{a great deal of} scholarly attention in recent years~\cite{lazer_science_2018,potthast2016clickbait,10.1145/3137597.3137600, pennycook2018prior, doi:10.1177/1529100612451018, vosoughi2018spread,altay2023survey}. \final{In this paper we} adapt Wardle et al.'s operationalization of misinformation as false information, false connections, and misleading content~\cite{wardle2018thinking}. Like Starbird et al., we understand misinformation as collaborative work within online crowds~\cite{10.1145/3359229}. \final{We explore criteria that can expose misinformation, and discuss the use of technology \added{to support} this exposure. Prior research has examined} what makes people prone to believe in misinformation~\cite{pennycook2020falls,Scheufele7662,altay2020if} and how the spread of misinformation can be studied on social media~\cite{flintham_falling_2018,pasquetto2020tackling,shu2019studying}. Notably, Pennycook et al. found that people's proneness to believe or share misinformation \final{is often not} because of motivated reasoning but because people do not pay attention~\cite{pennycook2019lazy,pennycook2021shifting}. Prior research also indicates that the role of individual rationality in the context of misinformation may be overstated, and that sharing of misinformation may be influenced by shared group-level narratives~\cite{sloman2018knowledge} and the social media context~\cite{pennycook2021shifting}. 

It is important \final{here} to consider different notions of ``sharing''~\cite{doi:10.1177/2056305115578135}. \added{In the social media context, many activities described as ``sharing'' are political and value-laden. Facebook, for instance, encourages peer-to-peer sharing to ``amass'' data about user activities~\cite{doi:10.1177/2056305115578135}. Considering the findings by Pennycook et al.~\cite{pennycook2019lazy,pennycook2021shifting}, Facebook's incentivization and users' tendency to ``share'' information without paying attention could, therefore, increase the misinformation problem.}

Considering the challenges posed by misinformation, simple ways of supporting users \final{in determining the reliability of a news website are needed. Shahid et al.'s recent study on misinformation in videos in India showed, for instance, that users are unwilling and lack the skills to make the assessment~\cite{10.1145/3491102.3517646}}. In a similar study on WhatsApp in India, Varanasi et al. found that simple and easy-to-process labels are preferable to more complex fact-checks~\cite{10.1145/3491102.3517646}. 

\added{As an important step towards simple and easy-to-process solutions, \final{we focus on the utility of reliability criteria for entire news websites. Qualitative work} by Jahanbakhsh et al. on why people trust individual news stories showed that this \final{trust} can be attributed to individual users' knowledge and experience~\cite{farnaz2021lightweight}. \final{Their identified} criteria include having firsthand knowledge\final{,} whether information is consistent with users' experience, whether other sources confirm claims, whether evidence is provided in the article, and whether the story is from a trusted source. This prior work informed us to explore criteria beyond individual claims, \final{since it can be hard for individuals to verify if they lack experience with news production}}.

\subsection{The Pursuit of Solutions Against Misinformation}

Researchers have explored a number of ways to support people \final{in identifying misinformation}. The most important prerequisite for this is that people can be persuaded to consistently and reliably change their beliefs \final{--- a possibility} shown in prior work~\cite{aldrich1989foreign,rabinowitz1989directional,wood2019elusive}. \final{Attempts} to address the problem of misinformation include checklists that help people consider the accuracy of the content they \final{want to share~\cite{farnaz2021lightweight,pennycook2021shifting} and reliability ratings for individual stories~\cite{10.1145/3415211}}.

Lazer et al.~\final{identify} two kinds of solutions against misinformation: \final{1)~changes} focused on the individual, e.g., empowering them to evaluate the misinformation, and \final{2)}~structural changes that prevent individuals from encountering misinformation~\cite{lazer_science_2018}. With this paper, we hope to contribute to both. \added{Reliability criteria can empower individuals to assess whether a website is reliable\final{, and o}nline platforms and media companies can use the criteria to add warnings to content and \final{to} remove certain websites in a transparent and explainable way.} 

\final{While a} large body of research on potential solutions examines the feasibility of using data mining and machine learning to automatically classify articles as misinformation~\cite{perez-rosas-etal-2018-automatic,wang-2017-liar,10.1145/3137597.3137600,10.1145/3305260,potthast2016clickbait,castillo2011information}, 
researchers in critical data studies highlight limitations in the quality and availability of \final{training data ~\cite{doi:10.1177/2053951719843310,doi:10.1177/20539517211017593}, noting} that since ML-based systems rely on correlations, their applicability is limited. For example, while data-driven approaches may help identify potential existing and debunked misinformation stories, their helpfulness for emerging stories is limited. 

One of the most promising non-technical solutions against misinformation that has emerged in recent years is the so-called lateral reading technique, pioneered by Wineburg and McGrew~\cite{wineburg2017lateral}. The technique is informed by the finding that professional fact-checkers make better decisions and need less time because they step out of a website and use additional information from other sources to make an informed assessment. For Wineburg and McGrew, \final{this lateral reading means reading less and learning more. Unlike those who stay on a specific page to evaluate a website's reliability, those who read laterally scan a website and then use other websites to judge the original site's credibility}. The effectiveness of lateral reading was shown in a large-scale, district-wide field study within high school government classes\final{, with results showing} that lateral reading can significantly improve high school students' ability to judge the credibility of digital content~\cite{wineburg2022lateral}. Our reliability criteria \final{support such lateral reading by helping end users decide what aspects to consider when using other websites to judge the reliability of a news website}.

\subsection{Reliability Criteria}

\final{In the early 2000s, Fogg et al. examined people's perception of the credibility of early websites~\cite{10.1145/365024.365037}. Elements that were seen to increase the perceived credibility included ``real-world feel,'' ``ease of use,'' and ``trustworthiness,'' among others. Perceived ``amateurism'' and ``commercial implications,'' on the other hand, were seen as factors that hurt the credibility of websites. Fogg et al. also provide an early example of studying the credibility of live websites with a focus on laypeople~\cite{10.1145/997078.997097}. Twenty years after these investigations, information environments have changed dramatically, especially news websites. These differences motivated us to focus on reliability criteria of current news websites.}

\final{Based on a literature review of keywords like ``reliability criteria'' and ``credibility criteria,'' as well as on recommendations by reviewers, we compiled a list of criteria from prior work aimed at helping users distinguish reliable from unreliable news websites. Many lists of criteria include no explanation of how the criteria were determined or of their empirical basis ~\cite{bradshaw2020sourcing,guess2020digital,newsguard_2022,TrustProject}. Two such lists are particularly noteworthy because they are widely used: the Facebook Tips~\cite{doi:10.1073/pnas.1920498117} and the News Guard criteria~\cite{newsguard_2022} (Table~\ref{tab:related_work_criteria}). Guess et al.~empirically showed the effectiveness of the Facebook Tips in the United States and India~\cite{doi:10.1073/pnas.1920498117}.} In the United States, the increase in discernment remained measurable after several weeks. \final{These tips were also included in a large-scale investigation by Arechar et al., who surveyed more than 33,000+ people in 16 countries across six continents, finding that combining the tips with subtle prompts to think about  accuracy improved the veracity of news people were willing to share across the globe~\cite{rand16countries}}.

\added{Other reliability criteria \final{without clear explanations} (i.e., the authors wrote that they selected certain criteria without also specifying their selection process) are \final{those} used by Bradshaw et al.~\cite{bradshaw2020sourcing} and the Trust Indicators by The Trust Project~\cite{TrustProject}. The criteria by Bradshaw et al.~\cite{bradshaw2020sourcing} include professionalism, i.e., whether best practices of professional journalism are followed; writing style; credibility, including whether false information or conspiracy theories are shared and corrections are provided; bias; and whether others' design and content strategies are imitated (``counterfeit''). The Trust Indicators by The Trust Project~\cite{TrustProject} include following journalistic best practices, having journalistic expertise, labeling types of content and reducing bias, referencing sources, \final{using} transparent reporting methods, locally sourcing stories, including diverse voices, and inviting and listening to feedback.}

\begin{table}
\small
  \caption{Reliability criteria from prior work on assessing websites. We include the Facebook Tips~\cite{doi:10.1073/pnas.1920498117} and the NewsGuard Criteria~\cite{newsguard_2022}. Both \final{sets of criteria} are widely used, and both lack empirical grounding.}~\label{tab:related_work_criteria}
  \Description{The left row of the table lists the ten Facebook Tips, the right column lists the NewsGuard Criteria. The ten Facebook Tips are: 1. Be skeptical of headlines, 2. Look closely at the URL, 3. Investigate the source, 4. Watch for unusual formatting, 5. Consider the photos, 6. Inspect the dates, 7. Check the evidence, 8. Look at other reports, 9. Is the story a joke, 10. Some stories are intentionally false. The NewsGuard criteria are: I. Does not repeatedly publish false content, II. Gathers and presents information responsibly, III. Regularly corrects or clarifies errors, IV. Handles the difference between news and opinion responsibly, V. Avoids deceptive headlines, VI. Website discloses ownership and financing, VII. Clearly labeling advertising, VIII. Reveals who's in charge, IX. The site provides names of content creators along with either contact or biographical information.}
\begin{tabular}{p{4.8cm}p{8.2cm}}
\toprule
\textbf{Facebook Tips~\cite{doi:10.1073/pnas.1920498117}}                        & \textbf{NewsGuard Criteria~\cite{newsguard_2022}}                                                                                \\
\midrule
\arrayrulecolor{lightgray}
1. Be skeptical of headlines            & I. Does not repeatedly publish false content                                                         \\ \hline
2. Look closely at the URL              & II. Gathers and presents information responsibly                                                      \\ \hline
3. Investigate the source               & III. Regularly corrects or clarifies errors                                                            \\ \hline
4. Watch for unusual formatting         & IV. Handles the difference between news and opinion responsibly                                       \\ \hline
5. Consider the photos                  & V. Avoids deceptive headlines                                                                        \\ \hline
6. Inspect the dates                    & VI. Website discloses ownership and financing                                                         \\ \hline
7. Check the evidence                   & VII. Clearly labeling advertising                                                                      \\ \hline
8. Look at other reports                & VIII. Reveals who's in charge                                                                           \\ \hline
9. Is the story a joke?                 & \multirow{2}{7cm}{IX. The site provides names of content creators along with either contact or biographical information} \\
10. Some stories are intentionally false &      \\
\arrayrulecolor{black}
\bottomrule
\end{tabular}
\end{table} 

\added{We identified two examples \final{in which} criteria are based on discussions with experts. Zhang et al.~\cite{zhang2018structured} used a subset of 16 credibility indicators \final{chosen} from over 100 proposed by representatives from journalism and fact-checking groups, research labs, social and annotation platforms, web standards, and others. They distinguish between content and context indicators. \final{\textit{Content indicators}} include title representativeness, ``clickbait'' titles, quotes from outside experts, citations of organizations and studies, calibration of confidence, logical fallacies, and inference (i.e., how correlation and causation are discussed). \final{\textit{Context indicators} include} originality, fact-checks, representative citations, the reputation of citations, the number of ads, the number of social calls, ``spammy'' ads, and the placement of ads. \final{The 16 chosen indicators require no subject matter domain knowledge, off-line investigation, or additional data gathering. Since they focused on articles, Zhang et al. also ignored indicators related to publishers, authors, and multimedia content.}}

\added{The second example of criteria based on experts' discussions are the Credibility Signals \final{proposed by the Credible Web Community Group}~\cite{W3C}. Here, again, criteria were discussed by experts. The following signals were selected after a review in 2020: the date a website \final{was first archived, and whether the site has a corrections policy, has won any awards, has won a Pulitzer Prize, or has won the Ramnath Goenka Excellence in Journalism Award.}}

\added{Considering the plethora of reliability criteria, it is important to understand \final{whether they are empirically grounded in the practices of end users and experts}. We identified only one example of reliability criteria based on a survey that studied this~\cite{10.1145/3415164}. Bhuiyan et al.~focused on individual news stories in the context of climate science and crowdsourcing news credibility assessment~\cite{10.1145/3415164}, \final{comparing} the news credibility assessments of students and crowd workers to those of three scientists and three professional journalists. Based on the responses of the six participants, they identified eight credibility criteria: accuracy, impartiality, completeness of coverage, originality and insight, credible evidence/grounding, publication reputation, professional practices and standards, and website aesthetics. Since these criteria hinge on the domain knowledge of individuals, we expand on them with a focus on news websites for the reasons explained in our introduction. Our study focuses on the whole website, a more extensive breadth of topics, and a much larger sample of consulted experts.}

\section{Methods}

To determine reliability criteria for news websites\final{, and to ground these criteria} in how journalism is done in practice and how \final{end users} think about the journalism they consume, we \added{combine elements of contextual inquiry~\cite{salazar2020contextual}, think aloud study~\cite{kuusela2000comparison}, and semi-structured interviews~\cite{kuniavsky2003observing}}. \final{It was necessary to combine these different elements, as contextual inquiry or participant observation alone would not have allowed us to understand the criteria participants use in practice, since these considerations are not observable. Conducting a survey was not feasible, since we wanted to capture participants' reactions to live websites with multiple articles and since our goal was to capture all criteria used in practice. Finally, structured interviews without the live websites and the think aloud method could have limited the insights we obtained, since participants would have had to rely on their memories about past interactions with unreliable websites.}

We studied two groups of people: professional journalists as experts, and politically diverse laypeople as end users. Recruiting from these two groups also allowed us to determine participants' political affiliation, which is important considering the complex role that partisanship plays in the assessment of news~\cite{10.1145/3491102.3502040,kahan2013ideology,pennycook2019lazy,pereira2023identity,farago2019we}. \final{For end users}, we recruited a representative sample of elected politicians in a Western democracy. This sampling ensured \final{participant diversity and a} full range of political views. In theory, the elected politicians in a representative democracy are representative of the voting body. In practice, people with higher education degrees are highly overrepresented in parliaments worldwide~\cite{erikson2019does}. We will reflect on this in the Limitations Section~\ref{sec:limitations}.

The target audience for \final{our news reliability criteria is the end user}. While \final{end users may lack} the expertise necessary to follow all of the practices experts use to assess a news site's credibility, experts’ practices are worth capturing in that 1) they represent a potential gold standard, 2) some of their practices may be teachable to \final{end users}, and 3) some of their practices may be accessible to \final{end users with the assistance of} future automated or socio-technical systems. Current \final{end users' practices in evaluating news sites' credibility are valuable for 1) capturing the range of current practices among those not professionally engaged in news production and 2) comparing these current practices to those of experts to better understand the gap that these reliability guidelines are intended to remediate.}

Since prior work \final{has shown} the importance of cultural and political differences, we identified appropriate field sites in two different countries --- the United States and Germany. We selected \final{experts} from the United States, a country with high concern about misinformation and high media polarization~\cite{newman_overview_2020,doi:10.1177/1940161219900126}, which these \final{experts had directly experienced. And we} selected a politically diverse set of \final{end users} that were representative of a federal state within Germany because that is where we had the best access to elected officials from across a wide political spectrum. While this prohibits us from drawing \textit{strong} conclusions about cultural differences or differences \final{in how end users and experts think about evaluating news sites, it allowed us to collect data across two different information and cultural landscapes, which we hope leads to a more generalized and therefore more useful set of criteria}. 

The research questions that drove our study design are:

\begin{itemize}
\item \textbf{RQ1:} What reliability ratings do \added{end users and experts} provide for news websites?
\item \textbf{RQ2:} What criteria do \added{end users and experts} apply to assess whether news websites are reliable?
\item \textbf{RQ3:} What differences between \added{end users and experts} can be observed?
\end{itemize}

\subsection{Procedure}
\label{sec:proc}

\begin{table}[]
\small
  \caption{ 
    \added{
    \textbf{Study Procedure:} Each participant reviewed three purported news websites following the structure in the table.
    }
  }~\label{tab:user_study_procedure}
  \Description{}
\begin{tabular}{p{13.5cm}}

\arrayrulecolor{black}
\toprule
\added{\textbf{Instructions \& Questions for the Tasks: Rate the Reliability of This Purported News Website.}}\\
\midrule
\added{1. First, participants were asked \textbf{whether they \final{had} personally used} the website.}\\
\added{2. Then \final{they} were asked to \textbf{freely explore the website} in their own browser at their own pace. We encouraged them to \textbf{describe the reasoning} behind their actions and decisions (\textbf{``think aloud''}~\cite{kuusela2000comparison}). Participants were allowed to read as many articles as they wanted, to visit all sections of the website\final{, and} to open other websites like search engines, social media websites, and encyclopedias.}\\
\added{3. While exploring the website, participants were regularly asked whether they \textbf{\final{felt} comfortable providing a rating}. If they answered affirmatively, they were asked about their rating. If not, they were allowed to continue browsing. Reviewing an individual website took around five minutes.}\\
\added{4. After participants \textbf{provided their rating} on a 5-point Likert scale, they were asked to \textbf{elaborate on their rationale for \final{the rating}}.}\\
\added{5. Finally, participants were asked whether \textbf{any information} that would have helped them make their decisions was \textbf{missing} during the study.}\\ 
\bottomrule
\arrayrulecolor{black}
\end{tabular}
\end{table}

In both the written and the oral briefings before the \added{study}, we explained that we had two goals: to understand how people decide whether \added{a website} is reliable and, as a consequence, to inform the design of software that can better support users in \final{making this determination}. The \added{sessions} were scheduled for 30 minutes, but some took considerably longer. 
At the start of each \added{session}, we collected \final{demographic information}. All \added{participants} were asked their age, gender, and highest completed educational degree. Politicians, \final{as representatives of end users}, were also \final{asked to identify their party affiliation, while journalists, as experts, were asked to identify the news organizations for which they currently work and worked for in the past}. 

During the \added{sessions}, each participant evaluated the reliability of three purported news websites. \added{Table~\ref{tab:user_study_procedure} describes the study procedure, the task, the instructions we gave, and the questions we asked. Participants freely explored each website in their own browsers, at their own pace, and while thinking aloud. \final{To make the study as realistic as possible, they interacted with live websites. \final{Since these websites were continuously adding content, the participants viewed different articles.}}}

This task, the \final{\textit{rating task}}, helped participants reflect on the reasons that make them consider a website reliable or not. 
For each purported news website, \final{we first asked participants whether they had personally used the site}. The answer options included ``Never,'' ``Rarely,'' ``Sometimes,'' ``Often,'' and ``Always.'' We then asked ``How do you think the reliability of the source should be rated?'' The participants provided their assessment on a 5-point scale, with the options ``Very unreliable,'' ``Unreliable,'' ``Partially reliable, partially unreliable,'' ``Reliable,'' and ``Very reliable.'' Participants were instructed to take as much time as they needed. Akin to iterative interrogative techniques like the five whys, we repeatedly asked participants \final{to explain the reasons for their rating}. Our goal was to determine the root reasons for the perceived reliability of a source~\cite{enwiki:1102936260}. \added{The study also included four short questions for a follow-up publication unrelated to reliability criteria.}

We transcribed the \added{audio recordings of the think aloud study and the semi-structured interviews\final{, and} analyzed the material using qualitative content analysis~\cite{mayring2021qualitative}, a method equivalent to thematic analysis~\cite{Braun2006,doi:10.1080/2159676X.2019.1628806}.} We followed axial coding principles~\cite{corbin2014basics}\final{, with the first author reading the texts multiple times and, moving back and forth through the material, doing an open coding of the \added{material}}. The German \added{material was} coded in German, and the \added{English} \added{material} in English. These codes were then discussed in weekly meetings with the second author\final{, which} helped us refine a set of well-defined codes. After clustering, splitting, and merging different codes into coherent groups, \added{we identified categories/themes and subcategories}\final{, and} improved them in weekly sessions until \final{we reached a} unanimous agreement.
The responsible authorities granted IRB-equivalent approval for the portion of the study conducted in Germany. The U.S.-based portion of the study was reviewed and accepted by the hosting institution's IRB. We obtained informed consent from all participants. 

\subsection{Selection of Purported News Sources}
Since \final{our participants were located in two different countries, we customized the news sites for each country. The politically diverse end users, all German speakers,} were presented with three German news sources, while the \final{experts were presented with three news sources in English \added{from the U.S}. All of the sources had been rated by reputable external news analysts as unreliable}. 

\final{For the \added{end user sessions} in Germany, we selected two sources from the 20 German domains most frequently visited by Facebook users~\cite{DVN/TDOAPG_2020}. The dataset is based on 40,000 URLs that Facebook verified via third-party fact-checks. Since the Facebook dataset only contained extreme cases \final{of content reported by users}, we also included one questionable source that is not extreme on the spectrum from very reliable to very unreliable, selecting a German website that has won a prize for alternative media but that was regularly criticized for publishing conspiracy theories about the COVID-19 virus and pandemic.}

For the \added{sessions} with \final{experts}, we sampled three English sources classified as unreliable \final{in the meta-ranking by Gruppi et al.~\cite{gruppi2020nelagt2019}, which combines reliability ratings} by Allsides, BuzzFeed, Media Bias / Fact Check, Open Sources, Pew Research Center, PolitiFact, and Wikipedia. Media Bias / Fact Check also provides a more fine-grained \final{website classification. We randomly selected two sources from those that Media Bias / Fact Check rated as ``Conspiracy Pseudoscience'' and one source from those rated ``Questionable Source.''}

\added{We used the described method to ensure the purported news sources were unreliable. The first author confirmed the content and tone of the different websites to be comparable to \final{those of unreliable websites, based on his five-year experience studying misinformation in German and English. Throughout the paper, we make cultural differences --- like the role of the German AfD --- transparent when they might influence the understanding of news reliability criteria.}}

\subsection{Participants}

\textit{Politicians as End Users.} We recruited a politically diverse group of 23 elected politicians from one of the sixteen state parliaments of the Federal Republic of Germany. \final{To do so, we contacted the president of the parliament through personal connections. The president authorized and endorsed our study and sent an invitation to parliament members}.

\final{For context, we briefly explain the political alignment of Germany's different parties, using Wikipedia's classification of these parties as a reference for their political alignment~\cite{enwiki:1030256103}. According to Wikipedia, the Christian Democratic Union~(CDU) is seen as center-right, the Free Democratic Party (FDP) as center to center-right, and the Social Democratic Party~(SPD) and Alliance '90/The Greens~(GN) as center-left. The party The Left~(LT) is classified as left-wing to far-left, and the so-called ``Alternative'' for Germany~(AfD) as far-right. At the time of our study, the SPD, The Greens, and The Left formed the state government}. The AfD is the first far-right party to win seats in the German Federal Parliament since the Nazis. \final{It is essential for readers without exposure to German politics to know that, at the time of \final{our study}, all other parties had resolutions not to cooperate with the AfD}.

We recruited 23 of the 84 elected representatives (27.4\%) of the Free Hanseatic City of Bremen. The distribution of party affiliation of the \final{end users is almost proportional to that of the last election}. We recruited seven \final{end users} from the CDU, six from the SPD, four from GN, one from the FDP, two from the AFD, and one from another far-right party called ``Citizens in Rage'' (German: ``Bürger in Wut''). 
To avoid identification of the \final{latter, we merged his data with that of the AfD members}. Both parties are far-right parties with similar policies.

Fifteen (65.2\%) of the \final{end users} were male, and eight (34.8\%) were female --- similar to the percentage of females in the state parliament of the Free Hanseatic City of Bremen (36.9\%). The mean age of the \final{end users} was 49.10 years old (SD=11.37); the oldest was 65, and the youngest was 30.
The sample of \added{end users} was diverse in terms of the highest education completed. All had completed the German equivalent of high school, four \final{had received} the German certificate of general qualification for university entrance but did not continue to higher education, two finished vocational/professional qualifications, two had a Bachelor's degree, 11 had a Master's degree, one had a PhD, and one had completed the state examination to be a lawyer.

\textit{Journalists as Experts.} \final{We used two distinct newspaper sampling strategies to recruit 20 journalists working for U.S.-based, English-language newspapers. The first was based on readership, looking at the ten most widely circulated newspapers in the U.S. in 2019~\cite{enwiki:1101330486}; we recruited seven experts} from these newspapers. We refer to them as JP --- i.e., JP1 is the first \final{expert} we recruited based on the popularity of their employer's newspaper. 
The second strategy was based on political alignment\final{, allowing us to include a politically diverse collection of news organizations. Using the meta-ranking of Gruppi et al.~\cite{gruppi2020nelagt2019}, we identified reliable newspapers classified as left, center, or right. We determined political alignment using Media Bias / Fact Check, an American fact-checking website that rates news sources. While these ratings have been criticized for not meeting the highest standards of rigor and objectivity,  Chołoniewski et al.~\cite{choloniewski2020calibrated} point out that, despite their imperfections,} they have been judged as ``accurate enough to be used as ground-truth for, e.g., media bias classifiers~\cite{10.1145/2487788.2488132,pennycook2018prior,pennycook2021shifting}, fake news studies~\cite{pennycook2019crowdsourcing,pennycook2019lazy,pennycook2020falls}, and automatic fact-checking systems~\cite{perez-rosas-etal-2018-automatic,potthast-etal-2018-stylometric,potthast2016clickbait}.''
We sampled seven reliable newspapers identified as politically left or center-left, seven politically in the center or ``least biased,'' and eight politically right or center-right. 

\added{For context, we briefly describe the political alignment of the major parties in the United States. Since the 1850s, the U.S. presidential election has been won by either the Democratic Party or the Republican Party. The Pew Research Center recognizes four Democrat groups --- Outsider Left (closest to center), Democratic Mainstays, Establishment Liberals, and Progressive Left (furthest from center), and four Republican-oriented groups --- Ambivalent Right (closest to center), Populist Right, Committed Conservatives, and Faith and Flag Conservatives (furthest from center)~\cite{doherty2021beyond}. Due to a lack of available resources on how to map different news sources to these fine-grained criteria, we relied on the criteria by Media Bias / Fact Check.}

For each newspaper selected \final{through our sampling scheme, we emailed the experts} listed as authors of the most recent news articles posted on the newspaper's official \final{Twitter}\footnote{Twitter has been rebranded to X. In the Methods and Results sections, we refer to Twitter because this was the name of the service at the time of the investigation. In the Discussion and Conclusion, we refer to X.} account. We ignored reporting on sports, celebrities, dining, and cooking.
In cases where \final{an email was not listed but a Twitter account} was open for direct messages, we contacted them via \final{Twitter} from the first author's personal account. 
After recruiting 11 \final{experts} who identified as male, we stopped recruiting \final{experts} with stereotypical male names in an attempt to recruit a more gender-balanced sample.

In the final sample, eleven of 20 \final{experts} identified as male and nine as female. The mean age was 35.5 (SD=10.27); the oldest \final{expert} was 61, and the youngest 21. Seven \final{experts} were recruited from among the Top 10 newspapers sampled by readership, and the remaining 13 from newspapers sampled by political stance. We recruited six \final{experts} from the left (we refer to them as JL), three from the center~(JC), and four from the right~(JR). All \added{sessions with \final{experts}} were conducted in English.
The \final{experts} in our sample were highly educated\final{; all but} one were working towards or had completed university degrees after finishing high school. One was currently working towards a Bachelor's while working as a journalist. Seven stated that a Bachelor's was their highest degree\final{; of these, four degrees} were in journalism. Twelve \final{experts} had a Master's degree, four in journalism.

The \final{experts} worked for a number of news outlets, including newspapers with international coverage and circulation like the New York Times, national papers like Newsday and USA Today, and regional papers like the Star Tribune (Minnesota), The Chicago Tribune, and Capitol News Illinois. Also represented were \added{online-only outlets like Salon, Slate, Vox, The Observer, and CQ RollCall}. ``Right center'' (rated by Media Bias / Fact Check as center-right) outlets included the Washington Examiner and HotAir. \added{Two or more participating \final{experts} represented} the following outlets in our sample: The Chicago Tribune, the Boston Globe, Foreign Policy, Media Matters, Newsday, Politico, and Vox.

\section{Results}
\label{sec:results}

Through a combination of contextual inquiry, think aloud study, and semi-structured interviews, we identified the reliability ratings our participants provided for different news websites (RQ1). Based on our observations of and their reflections on their process, we compiled a set of criteria participants used for arriving at those ratings (RQ2). We also examined the similarities and differences in how \final{end users and experts} approach this critical task (RQ3). 

We refer to the three German websites as DE1, DE2, and DE3\final{, and the three U.S.-based websites as EN1, EN2, and EN3}.

\subsection{Reliability Ratings Participants Gave to Purported News Websites (RQ1)}

\textit{Personal Usage.} \final{In general, the end users and experts} in our sample were not personally using the three purported news sources we asked them to assess. \final{Most of the German end users had never used the websites, with 22\% of end users at least rarely using DE1, 17\% using DE2, and 13\% using DE3}. 
One \final{end user (AFD2) stated that he frequently visited DE1, while another (SPD1) frequently used DE2}. 
Even fewer \final{experts reported that they personally used any of the sources we asked them to review. Seventeen} of the 20 had never used EN1, while the remaining 3 (JP7, JL2, JR2) stated that they used it rarely. None of the \final{experts} had personally used EN2 or EN3.

\textit{Reliability Ratings.} We found that \final{end users and experts} rarely rated the purported news sites as reliable. Only one \added{end user} (3\%) considered DE1 to be reliable~(AFD1), while more than half (52\%) rated \final{it} as unreliable (35\%) or very unreliable (17\%). Three \final{end users} (13\%) rated DE2 as reliable~(CDU4, GN3, SPD1), but almost six of ten (57\%) found \final{it} unreliable (22\%) or very unreliable (35\%). DE3 was never rated as reliable, and almost four out of five \final{end users} (78\%) perceived DE3 as unreliable (26\%) or very unreliable (52\%). \added{Table~\ref{tab:participants_and_ratings_de} in the Appendix provides all individual ratings, \final{along with the party of the end user rating the websites}}.

The ratings of the \final{experts} for the English sources \final{were} even more extreme along the spectrum from reliable to unreliable.
Only one \final{expert~(JP5)} rated EN1 as reliable (5\%), and only \final{one~(JC1)} rated EN2 as reliable. Nobody perceived EN3 as reliable.  
Fourteen of the 20 \final{experts} (70\%) rated EN1 to be either unreliable (25\%) or very unreliable (45\%), \final{17 of the 20 (85\%) considered EN2 either unreliable (30\%) or very unreliable (55\%), and 16 of the 20 (80\%) perceived EN3 as unreliable (20\%) or very unreliable (60\%). \added{A detailed overview of the individual ratings of each \final{expert} and information on how the \final{expert} was recruited can be found in the Appendix in Table~\ref{tab:participants_and_ratings_en}}}.

Overall, we found that the ratings in the U.S. \final{corresponded to the meta-ranking of reliability by Gruppi et al.~\cite{gruppi2020nelagt2019}, while the ratings in Germany were consistent with the crowd-sourced ratings validated by fact-checkers~\cite{DVN/TDOAPG_2020}}.
The two populations were not the same\final{, however,} in their recognition of the sources' unreliability: \final{experts} more frequently labeled sources as very unreliable or unreliable, while \final{end users} chose the ``partially reliable, partially unreliable'' option more frequently. 

\subsection{Criteria To Determine the Credibility of News Websites (RQ2)}

\begin{figure}[t]
    \centering
    \includegraphics[width=\linewidth]{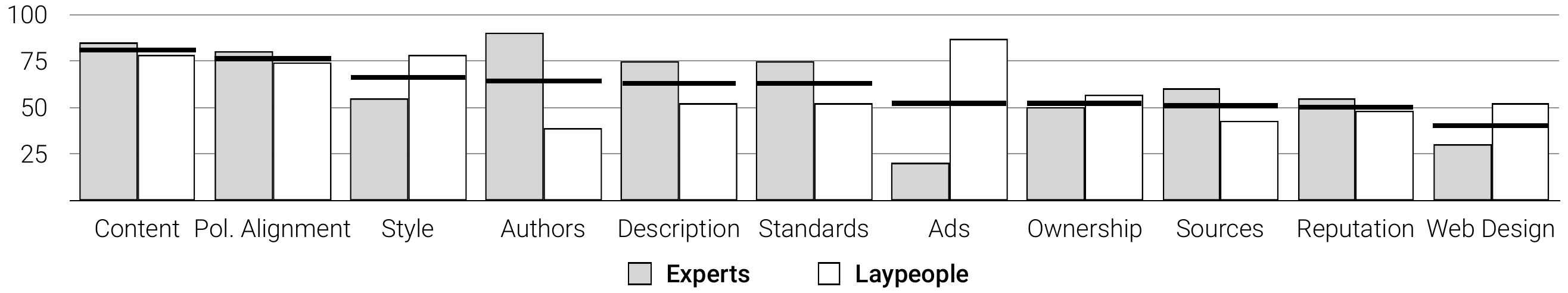}
      \caption{\added{Percentage of the 23 politically diverse \final{end users} from Germany and the 20 \final{experts} from the \final{U.S. who} used the eleven reliability criteria when assessing the reliability of a news source. The black line indicates the mean of both groups.}}~\label{fig:all_codes_small}
      \Description{The figure provides an overview of the different reliability criteria that we identified. The percentages indicate how many \final{end users} and \final{experts} used the criteria. Based on that ranking, the 1. content of the news websites is the most widely applied criterion, followed by the 2. political affiliation of a website, and the 3. writing style. Both \final{end useres and experts} also took the 4. authors and the 5. self-description of a website into account. Other important factors include whether a website follows 6. professional standards, the 7. advertisements on a site, and 8. the ownership of a website. The \added{participants} also considered 9. the sources used by a website, 10. the reputation of a website, and 11. the visual design and structure of a website.}
\end{figure}

The qualitative coding of the 43~\final{sessions yielded 11 criteria used by participants to determine the reliability of news websites. Figure~\ref{fig:all_codes_small} shows how many \final{end users and experts} used each criterion to evaluate the reliability of a purported news website; the black lines show the means of both groups}.
\final{In order of aggregated use across both end users and experts, the most widely applied criterion was the website's content, followed by its political alignment and its writing style. Both end users and experts also took into account the authors of the hosted articles and the website's self-description. Other important factors included whether a website followed professional standards, the site's advertisements, and its owners. Participants also considered the sources used by the website, the site's reputation, and its design}.

\final{We next} describe each criterion in detail, \final{distinguishing} between \final{end users and experts} to highlight \final{differences in how} they approached each criterion.

\subsection{Content}

\final{\textit{End Users.}} 
\final{Eighteen} of the 23 \final{end users}~(78\%) took the \final{articles'} content and topics \final{into} account when rating \final{a news source's} reliability. Important aspects in this regard include\final{d} whether more than one side of an issue \final{was} represented (FDP1, GN2), \final{and} how independent~(GN1, LT2), neutral~(CDU1), and objective~(GN1) the reporting \final{was}. \final{End users} considered how critically and thoughtfully something \final{was} discussed (FDP1), how controversial topics \final{were} presented~(GN1), and, more generally, the quality of the content~(SPD5). \final{End users} also found it problematic if opinions \final{were} disguised as news~(GN2) and \final{if} websites read like a blog~(SPD4). 

The content criterion connects to whether the information is compatible with people's knowledge and \final{whether they see it as} realistic (AFD1, AFD2, AFD3, CDU3, CDU6, LT2, SPD1, SPD2, SPD4, SPD5). CDU6, for instance, considered \final{whether she had} heard about an incident from different people: ``When everyone reports the same thing, then it must be real.'' SPD5 described this as a plausibility check. SPD1 used topics he is knowledgeable about to judge \final{source} reliability. \final{Topics mentioned by end users} included COVID-19~(LT1, SPD3) and mandatory vaccinations (FDP1), anti-Israeli or antisemitic attitudes~(GN2, SPD4, SPD4), \final{and} right-wing conspiracy theories~(GN1, LT1). FDP1 found it unusual that one of the websites did not cover topics like politics, sports, and society.

GN2 distinguished between misinformation that is completely made up, e.g., an invented incident involving foreigners and crimes, and articles that \final{reinforce} stereotypes and provide certain opinions in one-sided accounts. SPD3 included whether facts are twisted as an \added{essential} criterion. The \final{end users looked at whether an article was} from a news website or a blog~(CDU3, GN2, GN4, LT1, SPD3, SPD4), and whether it \final{was an op-ed or a source} spreading propaganda~(LT1). The amount of content was also an important criterion. SPD4 argued that if a source would produce that much content, ``then one would have heard from this source already''.

\final{End users also commented} specifically on whether websites were perceived as spreading conspiracies or promoting conspiratorial thinking, including right-wing conspiracy ideologies or propaganda~(GN1, GN4), especially antisemitic and Anti-Israeli positions~(SPD4).

\textit{Experts.}
\final{Seventeen} of the 20 \final{experts}~(85\%) recognized the content a news website publishes and the topics covered as important criteria. JP5\final{, for instance,} commented on the selection of stories, \final{expecting} that important current news stories would be covered. JL5 \final{considered} whether a website covered a story in the same way \final{as} a reliable source and whether the facts stated in the article fit what she \final{knew} from other sources. 

When assessing the reliability of news sources, the \final{experts} took the type of website into account. \final{They} distinguished between sources that provide breaking news~(JP5, JC1, JR1) and websites that do editorial, commentary, or opinion writing~(JP1, JP5, JC2, JC3, JR1, JR4) or provide analyses~(JR4). \final{Experts} also differentiated between traditional or mainstream media sites with editorial processes~(JC1, JC2, JR3), curated websites that collect articles from other sources~(JL1, JL3), and blogs~(JP1, JC3). JC2 and JP2 \final{expected a website to} clearly indicate that it is providing an opinion. The \final{experts} also took into account whether the content is actually relevant for the stated target group, e.g., military veterans~(JR1). JR1 \final{believed that news sources must be neutral}. JL1 believe\final{d} that reliable websites give news free of personal opinions. While he believe\final{d} that there are \final{stories that have} four or five sides, there are many \final{others with} only one side: the earth is \added{round}, man did land on the moon, and the Holocaust happened. JL1 reviewed all articles on one website and found that ``these are all lies --- I mean, these aren't even good lies.'' For JP2 and JP3, opinions need to be labeled as such. JP2 felt that opinion and news ``bled into each [other]'' in one of the sources. \final{Experts} criticized websites they perceived as passing opinions as news~(JP2, JP3, JL2, JL4, JL6). 

JL6 commented that not everything written by one of the websites was inaccurate but that \final{he was still skeptical about the sites}. JC1 wanted to understand whether a news source \final{had} an accurate grasp of the subjects they \final{reported} on. To quickly check a website, she \final{focused} on statistics she \final{could} easily cross-check, e.g., via government websites. Considering the breadth of content on one of the sites, JP2 wondered whether a source \final{had} reporters in all countries covered or, if not, where the information \final{came} from. JL6 found that ideologically driven sites like the one he was reviewing might write something factually correct, but \final{without context. He} argued that such sites do a lot of ``framing by omission,'' which \final{made} him skeptical.

The \final{experts} took several other content-related factors into account, \final{including} whether a well-established news outlet would cover a story~(JP2), \final{and a website's} categories and how they \final{were} determined~(JP7, JL2). JL2 perceived news sites with categories like ``Disloyal Military,'' ``Progressive Fascists,'' and ``Racist Mayors'' as unreliable. JL6 \final{doubted the reliability of the reporters'} fact-finding and whether \final{it was} based on original research, \final{and JP7 found the information highlighted on the sites to be} false or delivered with a ``spin'' (JP7). 

Several \final{experts~(JP3, JP5, JP6, JL2, JL6, JR2, JR3)} commented on the \final{conspiracy-related} and antisemitic content of one of the websites, \final{noting} conspiracies like the so-called ``great reset''~(JP7, JR2) and the ``deep state''~(JP5) as well as misinformation about vaccines~(JP6). JR2, writing for a center-right news source, described these topics as connections to ``the crazier sides of the conservative ecosystem.'' \final{Experts} noticed a general conspiratorial feeling~(JP5) or pointed out specific topics like Freemansory~(JR2, JR3) and Satanism~(JL6). JP1 and JP2 \final{recognized} right-wing tropes. \final{Specific} content made \final{experts} suspicious, e.g., alleged plans for gun permits based on political views~(JR3), allegations that U.S.~President Biden pays to promote atheism worldwide~(JR1), \final{the alleged involvement of NATO} in Sri Lanka's current financial crisis~(JC1), and Ukraine ``becoming the new Israel''~(JL1). JP2 felt that a story on Russia and China announcing a new global reserve currency \final{preyed on fears held by} ``evangelical Christian circles'' and people on the far-right end of the American political spectrum.

\subsection{Political Alignment}

\final{\textit{End Users.}} 
The political alignment of a news source was taken into account by 17 of the 23 \added{politically diverse \final{end users}}~(74\%), \final{who found} it especially relevant if a source was perceived to be affiliated with the political right~(AFD3, GN1, GN2, GN4, SPD1, SPD3, SPD4, SPD6) \final{--- in particular, the far-right AfD}~(CDU1, CDU4, FDP1, GN1, GN4, LT1, LT2, SPD6). \final{End users} also commented on a news site's alignment with the political left~(GN1, LT4, SPD3, SPD1, SPD6) and \final{with} political parties in general~(CDU4, GN4). Other topics related to political alignment \final{included} the relationship with and portrayal of Russia~(AFD1, AFD3, CDU5, CDU7, FDP1, SPD3); \final{SPD3,} for example, criticized one of the sources for blindly supporting Russian politics, \final{and CDU7 found} ads for any party problematic. More generally, \final{end users} highlighted the importance of \final{a website's neutrality}~(CDU4, CDU7), \final{looked at} who was sharing a source --- \final{with CDU1 noting} that a certain source was shared by people critical of immigration --- \final{and considered the} political alignment of those cited or interviewed within the articles ~(SPD3).

\final{\textit{Experts.}} 
\final{Seventeen} of the 20 \final{experts}~(80\%) used the political alignment of a source to determine its reliability. JR1 argued that news websites should be about disseminating information free of bias, slant, and political mission. JL1 argued that a good journalist is a disinterested third-party observer, \final{elaborating} ``The journalists that are far-right: not journalists! The journalists that are far-left: not journalists. The journalists who always try to be in the middle: may not be journalists either.'' JP5 \final{negatively} perceived one of the sources as having an agenda and publishing opinionated content. \final{The perceived political alignment} included specific groups like the political right-wing or far-right~(JP1, JP2, JP3, JP4, JP7, JL1, JL2, JL3, JL5, JR2), conservatives~(JP2, JP5, JL3, JL4, JR1, JR3), Islamophobes~(JP1, JP2, JP3, JP7, JL1, JC3, JR1), antisemites~(JP3, JL2, JL5, JL6), and pro-Russians~(JP3, JP6, JL5). \final{Experts} \added{considered} \final{whether} sources positioned themselves against progressives~(JP4, JP5, JP6, JL1, JL4, JL5, JR1, JR2) \final{or} the left~(JP1, JR2, JP7, JC3), \final{and} referred to fringe content or extremism in general~(JP3) or \final{to} other groups~(JP2, JL1, JL5). \final{JR1 saw} \added{political alignment} as problematic, \final{saying} newspapers should not have a political position. \final{Experts} also viewed patriotism and statements that a source is ``100\% pro-USA'' as problematic~(JP4, JP6). JL2\final{, however,} stated that just because a source has a different ideological perspective does not mean it is not useful or \added{is automatically unreliable}. JR3 distinguished between sources that want to give people reliable news and sources that support certain political causes, but how she \final{made} this distinction was not captured. \final{Experts} also discussed how the ideological lens of a website shapes its motivations~(JL2, JL3), \final{with JL2 finding that one of the sources was not even pretending to present non-biased news but was instead} responding to the news of the day with a very specific ideological lens. 

\final{Experts} described a number of ways in which they determined the alignment of websites and authors. The Southern Poverty Law Center, \final{for example,} an organization that monitors extremism, informed the perception of a website's political affiliation~(JP1, JP3, JL5). \final{Experts also} recognized political alignment through tropes \final{they} associated with political parties or partisans~(JP1, JP4). \final{Experts} recognized a right-wing political bent~(JP1, JP2), e.g., in stories about crime in big cities with progressive mayors, \final{considered a familiar tactic by right-wing media~(JP1)}.  

\subsection{Writing Style}

\final{\textit{End Users.}} 
\final{Eighteen} of the 23 \final{end users}~(78\%) used the writing style of a source as a criterion \final{to judge reliability, describing} wording as ``lurid''~(AFD3, FDP1, SPD2, SPD3, SPD4, SPD6), ``populist''~(CDU4, CDU5, LT1), ``sensationalist''~(CDU5, CDU6, CDU7, GN3, SPD4), and ``propaganda''~(GN4, LT1). \final{They} also perceived the websites as ``polemic''~(CDU2), ``manipulative''~(GN3), ``heretic''~(CDU5), ``clickbait''~(SPD2), and ``right-wing agitation''~(GN4). AFD3 and GN4 argued that \final{a headline's} writing style can sometimes \final{be} sufficient evidence \final{of} ``fake news'' or political bias, \final{citing examples such as} ``Media go to war against Russia''~(AFD3, CDU2) and ``Compulsory masks are child abuse''~(GN2, LT1). For GN4, the keywords used by the political right would lead him to judge a source as ``very unreliable,'' \final{while SPD5 also took into account the terminology's precision and subtlety}.

GN3 commented on specific techniques like exaggeration, talking down, and trivialization. AFD3 criticized allusions like ``Russia's Red Lines and Who's Really Escalating.'' AFD3 and CDU2 \final{commented} on the practice of posing questions, \final{with CDU2 citing} the question ``Is this what modern synagogues look like?'' as an example that made her cautious. 

For SPD4, certain words \final{reminded} him of other far-right conspiracy websites. \added{GN3 said the term ``lust for war'' alone stopped her from reading}. \final{Similar} examples included ``vaccination enthusiast''~(GN1, GN4), ``lab rats''~(GN1), ``war-mongering''~(SPD4), and ``attitude journalism''~(GN4). LT2 highlighted the term ``genocide of Russians'' as biased. CDU7 criticized terms like ``resistance to [the vaccination] terror,'' which she considered factually wrong. \final{Several end users} compared the wording of the websites to a commonly criticized German tabloid paper~(GN3, SPD4, CDU6, CDU7). LT1 \final{found} the wording of one source so populist that it \final{did} not allow an objective political debate.

\final{\textit{Experts.}} 
\final{Eleven} of the 20 \final{experts}~(55\%) used writing style as a criterion. \final{They} took tone into account~(JP2), evaluated \final{the language's objectivity and neutrality~(JP1, JR1), and} criticized the writing for not being clear~(JP6, JC2), e.g., when a source did not indicate whether authors spoke to a source or were citing a secondary source~(JC2). JP6 criticized a source when language dehumanized and villanized people, \final{and JC1 looked at whether the writing went} straight to the point or was ``rambling''~(JC1).

\final{Experts described the} language of the reviewed articles as ``distorted''~(JR4), ``inflammatory''~(JP1), ``biased''~(JC3), ``provocative''~(JC2), ``hyperbolic''~(JR2, JR4), ``overwrought''~(JR4), and ``clickbait''~(JC2). \final{JR4 saw the} hyperbole as enforcing a certain point of view. \final{Experts} also negatively commented on headlines they perceived \final{to have} an ideological slant~(JL2, JR3). For JL6, the editorializing of the headlines was \final{the reason} he considered one  source unreliable. \final{JP2 described the} tone of one website as right-leaning. \final{Experts} criticized the use of adjectives, e.g., in ``delusional Biden''~(JC1), and \final{of} superlatives, e.g., the allegations that U.S. \final{P}resident Biden ``destroyed the greatest military in the world''~(JP1, JP2, JL3) or that somebody made the ``worst pro-abortion argument ever''~(JP5).

\final{Experts} criticized statements that \final{were} not objective~(JP4, JP6, JR2), \final{and} criticized the source if a headline did not tell the story~(JR3). \final{They} also commented on punctuation~(JC1, JR3) --- e.g., the use of question marks~(JR3) and exclamation marks~(JL6, JC1) --- \final{and on} capitalization~(JR1) and ellipses~(JC1), which \final{were} seen as diverging from standard journalitic practices~(JC1). JC1 argued that exclamation marks should be avoided in journalism because they ``imply an emotion.'' 
JP2 interpreted the headline "We did it, Joe" as a snarky statement that held U.S. President Biden responsible for the recession. The wording of the headline led her to believe that the source was biased.

The \final{experts} also highlighted vocabulary~(JP4, JR2, JR3), \final{connecting u}nreliability to certain \final{terms} like ``progressive fascists''~(JP1, JP6, JR2), ``disloyal military''~(JP1, JR2), ``tyranny of woke intentions''~(JR1), ``great reset''~(JR2), \final{and} ``Zio-globalist handlers''~(JP1), and \final{to} neologisms like ``Fauci-baric''~(JR1). \final{Experts perceived terms such as} ``manifesto'' as ``loaded'' and ``politicized''~(JP4), \final{and were also triggered by terms such as} ``Freemasonry''~(JR2), ``fascist''~(JP6), and ``racist''~(JP6). \final{Experts} \final{noted} tropes of political parties and political partisans~(JP4) or conspiracies~(GN1). JR1 criticized \final{how one source} referred to paid advertisements as ``the Information,'' \final{attempting to cover up that the article was}  actually a paid advertisement.

\subsection{Authors}

\final{\textit{End Users.}} 
\final{O}nly 9 of 23 \final{end users}~(39\%) took the authors of articles on a news website into account. \final{These end users criticized websites on which authorship} was not clear ~(CDU6) or  information about the author was hard to find~(SPD4). CDU3 connected the question of who is writing to \final{their} motivations. \final{LT1 looked at} whether multiple journalists \final{we}re responsible for the content~(LT1), \final{and noted whether a journalist's background} and photo were provided~(LT1). \final{End users} also took into account \final{whether} an author work\final{ed} for other, more widely known publications~(LT1), \final{and} considered whether they were familiar with certain authors~(LT1), who the guest authors were~(SPD6), and whether freelancers were employed~(CDU3). 
In addition, the\final{y} viewed \final{authors} favorably if \final{readers} \added{could} provide input and leads via email~(SPD6). 

\final{\textit{Experts.}} 
\final{The author of a news story was the criterion used by most experts. Eighteen of the 20 experts~(90\%) used it to judge a source's reliability}. The byline was seen as crucial~(JP1, JP2, JP4, JP6, JP7, JL1, JC2, JC3), \final{with experts using} information about the authors to understand where they had worked~(JP1, JL3, JC2), their ``beat''~(JC2), and whether they \final{were} professional writers~(JC2). For JL1, it was important to ``know why [the authors] are saying what they're saying.'' \final{Experts also} thought it important to know the \final{site's} editors and owners ~(JR4) and how \final{the authors were} funded~(JL4). \final{They} examined whether \final{an author worked} for a ``known organization,'' whether the work \final{was} funded by a known organization~(JC1), and whether \final{an author was affiliated with} a university. \final{They perceived an article} as reliable if more than one \final{author} was involved~(JP2, JC3) and criticized sources centered around one personality~(JL5, JR1). \final{Experts were} put off by the profile of one of the authors~(JP6, JC3), because it showed a ``random picture of a baby smoking a cigarette.''

\final{Experts} criticized \final{a} source if no \final{author} information \final{(e.g., a biography) was} available~(JP6, JC2, JC3) and if the \final{author's} qualifications \final{were} not presented at the top or bottom of a page~(JP1). JP2 found it important \final{to be} able to verify that somebody is a real person with an education and background in journalism. JL1 appreciated knowing where articles were curated from. \final{Articles written} anonymously or under a pseudonym \final{were} seen as problematic~(JP1, JP2, JP4, JP6, JP7, JL4, JL6, JC2, JC3) because it \final{was} difficult to judge \final{the credibility of an} individual author and their previous record~(JP1).

\final{Experts} consulted other sources to understand who the authors were~(JP1, JL1, JL3, JC1), \final{with one using the ``About'' page}~(JP4), while another consulted the Wikipedia article about an author~(JP1), \final{also noting whether} the Wikipedia article was ``comprehensive'' and ``well-cited''~(JP1). \final{Experts} also considered how institutions they trusted --- like the Southern Poverty Law Center --- describe\final{d} the authors~(JP1), \final{and whether} the content posted by an author was a repost from another website, e.g., the author's personal blog~(JP1). Other criteria included how frequently something was reposted~(JP1) and whether the authors had professional-looking email addresses from the website \final{for which} they were writing~(JC3).

\subsection{Self-Description}

\final{\textit{End Users.}} 
\final{Twelve} of 23 \final{end users}~(52\%) evaluated a source's self-description \final{in determining reliability, with SPD6 noting that} it is worth reading self-descriptions to understand what a website describes as its mission. \final{Some found it problematic when} sources described themselves as ``critical''~(CDU6, FDP1, SPD2), ``dissenting''~(CDU6), or ``independent media''~(CDU1, GN1), \final{and perceived slogans} like ``Truth and Tradition'' as dubious and untrustworthy (FDP1, GN4, LT1, SPD2). \final{For FDP1 and GN4, explicit statements that a source was telling} the truth made them skeptical. The slogan ``The critical website'' was interpreted as implying ``The others are not as critical as we are. We claim to be more critical.''~(CDU2), which was seen as problematic~(SPD2). GN1 and GN2 commented on the slogan ``For those who still have their own thoughts;'' for GN1, it insinuated that the \final{source's} authors and readership were the only ones who still thought freely, \final{a \added{self-description} that} made him very skeptical. GN4 criticized sources that discredit\final{ed} others to \added{make themselves appear better}, \final{and end users also commented on the} name of a source itself~(AFD2, CDU6).

\final{\textit{Experts.}} 
\final{Fifteen} of 20 \final{experts}~(75\%) used \final{a website's} self-description to determine reliability, \final{seeing it} as a way to understand editorial processes~(JP6, JP7, JL4, JR1, JR3), the editorial line or mission~(JP4, JL4, JL6, JC2, JR1, JR3), and the \final{website's} ownership and independence~(JL3, JL4, JL6, JC3). \final{Experts} used the description to understand the goals and motives \final{behind the site}~(JL3, JR3), ``who these people are''~(JP4, JP7), what \final{the website stood} for~(JR3), and whether \final{it had} a political alignment~(JL6). JR1 felt that a mission statement should be about delivering news, JP4 expected a commitment to journalistic objectivity, \final{and} JR1 argued that the mission statement should be free of bias, slant, and a political mission. JP4 and JC2 \final{noted} which corporate websites and charity groups \final{were} mentioned. \final{Experts} also commented on the opinions and perspectives they found in the self-description~(JC3, JR2), \final{using them to understand a source's} target group~(JL3, JR3). Statements deemed \final{by experts} as not objective frequently related to certain political groups, like the political left~(JC3, JR2) and Islam~(JC3, JR1, JR2), as well as \final{to} claims that a source was ``100\% pro-USA''~(JP4, JP6) or defending ``free society''~(JR1).

Unlike \final{end users}, \final{experts} rarely commented on the name of a source. Some commented on slogans~(JL3, JL5, JR1), like \final{those saying} a source was serving the clandestine community or veterans~(JL3), or \final{on} a banner that said that ``inside every progressive is a totalitarian screaming to get out''~(JP5, JL4, JL5, JR1, JR2). 

Surprisingly, \final{with the exception of JL3, JL6, and JC3, experts rarely commented on} the reliability of \final{the} self-disclosed information. JL3 stated that ``there's no way for me to know that for sure,'' JC3 commented ``I guess I don't know if it's real,'' \final{and} JL6 believed that \final{even when} the ``About'' page includes self-flattering things about a source, \final{it} can provide ``some insight as to how they perceive themselves.''

\subsection{Professional Standards for Journalists}

\final{\textit{End Users.}} 
\final{Eleven} of the 23 \final{end users}~(48\%) used whether a source followed professional standards as a criterion \final{in judging reliability, considering} how ``independent''~(GN1, LT1), ``objective''~(GN1), or ``critical'' (FDP1) the \final{website's} perspective was. They also examined whether multiple sides of a story \final{were} discussed (FDP1, GN2). SPD5, for instance, highlighted the importance of a ``journalistic quality of the preparation and presentation of content.'' SPD4 argued that one website he reviewed felt like a blog, \final{and} found this problematic because he believed that certain journalistic and self-regulatory aspects do not apply to blogs. In relation to this criterion, \final{end users} took into account whether they found something to be a ``journalistic contribution''~(GN1) and whether a website was perceived as hiding or selling opinions in the reporting~(GN2, SPD5). For LT1, this meant ``how close'' the writers of a website \final{were} to the subjects. SPD6 also \final{considered} whether a source was a member of journalistic associations which subscribe to certain ethical professional standards.

\final{\textit{Experts.}} 
\final{Thirteen} of the 20 \final{experts}~(75\%) explicitly commented on professional standards, the neutrality of a news source, and the clear distinction between opinions and facts~(JP1, JP4, JP7, JL1, JL2, JC2, JR1). \final{They criticized the reflection of} \added{a journalist's} personal opinions or political ideology in the writing~(JP2, JP4, JL1, JC3). \final{Experts} also commented on the bias of one source they reviewed~(JL4, JC3). JL1 \final{believed that} a good reporter assesses information individually and does not take sides, \final{and a} reliable source should provide news without a viewpoint. JP7 \final{believed that} journalism should show a set of facts. JP4 named striving for journalistic objectivity as the ``normal standard.'' \final{For JP7, it was} a red flag if only one side of the political spectrum \final{was} represented and if the content \final{was} portrayed through a political lens; \final{he} perceived one of the sources as adding a spin to the content. JC2 argued that a website has to indicate if it provides opinions \final{only} for a particular segment of the political spectrum. \final{He saw claims of patriotism or pro-USA sentiment as indications of} not being critical enough of the U.S.~government~(JP4).

JL1 highlighted the importance of factual, vetted information free of opinions and preconceived notions, \final{and} stressed that opinions should be clearly labeled. \final{It} was important to \final{experts} like JL1 and JR1 that knowledgeable copy editors check the writing. JL1 argued that, when gathering information, it is important to always attack it ``fresh.'' As somebody working in the White House, he perceive\final{d} his job as giving ``every president shit'' and questioning power. In his personal experience, he found that ``because I went after Trump, I was [perceived as] far-left --- and now, because I go after Biden, I'm [perceived as] far-right.''

One news source wrote that editors and writers \final{were} not employees or contractors of the site and \final{were} not paid\final{; t}his was seen as a red flag by a number of professional \final{experts}~(JP7, JL4, JC2, JR1, JR3), \final{who linked it} to a lack of editorial control and vetting~(JP7, JC2, JR1). \final{Experts found it problematic that there were no editors checking} content and claims in the articles for veracity, and no copyediting on the writing itself~(JC2, JR3, JP7, JL4).

\subsection{Advertisements}

\final{\textit{End Users.}} 
The criterion that \final{end users used} most frequently to assess \final{a source's} reliability was advertising\final{, with} 20 of the 23 \final{end users}~(87\%) \final{referring} to ads when explaining their reasoning. They \final{commented} frequently on the number~(CDU3, CDU4, FDP1, SPD4, SPD6), content~(CDU6, FDP1, GN3, SPD4), presentation~(FDP1, GN3, SPD3), and size~(CDU4, SPD4) of ads, and on apparent disconnects between a \final{site's} content and its ads (FDP1, GN3). Ads were considered ``dubious''~(CDU1, CDU6), e.g., if they featured diet pills~(CDU6) or ``tricks'' \final{to} repair eyesight (FDP1). CDU3 perceived \final{a large number of} ads as indicative of financial dependence and political bias. \final{End users} also commented on \added{how} ads were integrated into \final{a website's} design~(CDU4, FDP1). FDP1 criticized \final{how ads were not labeled as such,} and layouts that made it hard to distinguish between content and ads.

\final{End users} considered whose ads were displayed~(CDU1, FDP1, GN1, GN4, LT2). One of the websites evaluated included ads for a book by politicians from the far-right AfD party\final{; this was criticized by end users from} across the political spectrum~(CDU1, CDU4, CDU7, FDP1, GN1, GN2, GN4, LT2, SPD3). CDU7 \final{believed} that such ads limit\final{ed} the \final{source's neutrality, saying} she would also criticize \final{an} ad if \final{the} book was from her party. For GN1, ads for the AfD gave him an ``absolutely clear'' impression of the source because he has a clear opinion of the AfD as a party. LT1, too, argued that since the AfD is not a ``respectable'' party, a website advertising a book related to the AfD is \final{not} ``respectable.'' For CDU1 and LT2, the \final{AfD} ad meant that the website was ``obviously far-right.'' For them, \final{a} populist ad was reason alone to leave a site and an indication that a source \final{was} related to the political right. The same \final{was} true for far-right publishers~(GN2, LT1).

\final{\textit{Experts.}} 
Only 4 of the 20 \final{experts}~(20\%) referred to advertisements when making their assessment\final{s}. The \final{experts} required websites to clearly distinguish between ads and content~(JP7, JR1), \final{criticizing sites if an ad was} the first thing shown~(JL1) and if \final{the ad looked} like an article~(JC3). The \final{appearance of} one ad  was seen as \added{an} indication of a ``typical unreliable website''~(JL1). Another important criterion for \final{experts} was whether major ad networks cooperated with a website~(JC3, JR1). JC3 argued that ``if it gets to the point where the ad networks are like --- we don't want to work with you --- that kind of raises a red flag." JR1 noticed that one of the sources described itself as paid advertisers \final{who publish favorable information, and saw this} as unreliable.

\subsection{Ownership}

\final{\textit{End Users.}} 
\final{A website's ownership was used as a criterion by} 13 of the 23 \final{end users}~(57\%), \final{who asked questions such as} ``Who is behind it or who owns it?''~(AFD1, CDU5, GN2, SPD4, SPD6), ``What interests are behind it?''~(CDU5), and ``Who finances it?''~(CDU5). SPD6 \final{looked at} what type of company \final{ran the} website, whether the \final{company's goal was} to make a profit, and whether \final{the} website \final{was} financed through ads \final{or} subscriptions or \final{was} fee-financed under public law. SPD2 \final{considered} whether a publishing company was behind a site and whether she \final{saw} this company \final{as} trustworthy. \final{End users} like CDU6 stated that they regularly look at the \textit{Impressum} of a website. In Germany and other German-speaking countries, the \textit{Impressum} is a legally mandated statement of \final{a document's} ownership and authorship ~\cite{enwiki:1092341470}. This information was often challenging to find~(CDU5, CDU6, LT1, SPD4), \final{a fact criticized by}~CDU5. \final{End users} also took into account whether they \added{were} familiar with the \final{website's} publisher~(SPD3).

\final{\textit{Experts.}} 
\final{Ten} of the 20 \final{experts}~(50\%) took ownership into account when assessing \final{a website's reliability, finding} it helpful to understand whether a website \final{was} funded by a known organization~(JC1) and whether \final{that organization had} an agenda~(JL4). \final{They wanted} to understand the owner, their motivations, and their prior work~(JL3, JR1). \final{In} one source, the owner was clearly named, \final{and this was noted by several experts} (JP1, JP3, JL3, JL4, JR1, JR3). JP3 had professional interactions with \final{this} owner, and did not hold him in high regard. \final{Experts} researched the \final{website owners on} Wikipedia, Google, and \final{Twitter}~(JP1, JP3, JR3, JP4, JL4), \final{taking} into account what \final{was reported by} institutions or sources they trust\final{ed}~(JP1) and who the owner associated with on \final{Twitter}~(JR3).  

\final{One} important criterion was how a source was financed~(JR1). A website \added{claiming} to be independent was perceived positively because ``they are not owned by any billionaire''~(JL3). However, not taking money from outside media sources, the government, and sponsors was also seen as unreliable by one of the \final{experts}, JR3, who trust\final{ed} traditional media partners, \final{and} also mentioned the board of an organization. JP4 said he would try to access IRS forms to see who the funders \final{and} officers \final{were}, and how the money \final{was} spent.

\subsection{Sources}

\final{\textit{End Users.}} 
\final{Ten} of the 23 \final{end users}~(43\%) \final{noted the sources that were} cited and linked and \final{the type of} supporting evidence provided. \final{End users believed that} reliable sources should transparently document and cite evidence for reports, studies, and comments~(FDP1, LT2). Examples of reliable sources \final{mentioned by end users} included public broadcasting, newspapers of record, and public authorities. SPD2 highlighted the importance of scientific findings, e.g., regarding the COVID-19 pandemic. \final{Other criteria included} whether more than one source was cited and the critical engagement with these sources. Some \final{end users} also appreciated \final{the citation of sources} they considered reliable~(LT1, SPD2).

\final{\textit{Experts.}} 
\final{Twelve} of the 20 \final{experts}~(60\%) referred to the sources used by a website \final{when evaluating its reliability. They} highlighted the importance of original reporting~(JP5, JP7, JL4) --- in particular, first-hand accounts~(JP3, JP5, JL1), interviews, \final{and} direct quotes~(JP1, JP3, JL1, JL3), especially with named sources~(JL1, JC1). JL1, for instance, argued that the only way to determine if a news story is real is to witness the event or to talk to somebody who witnessed \final{it. Experts also mentioned direct consultations with} domain experts~(JP1, JP2, JL1, JC3) \final{and the use of} wire services~(JP1, JP5, JL1, JR3) and other newspapers~(JP3, JP7, JL1, JR3), especially those \final{which} are reliable or centrist~(JP3). \final{They} considered where a source \final{obtained} its material ~(JR1) and how the information \final{was} vetted~(JL1). When vetting sources, the \final{experts} \added{considered} who the sources \final{were}~(JL4) and what connections the sources \final{had}, e.g., whether they \final{were} affiliated with a government or a certain ``oligarch''~(JP5). The \final{experts} referred to authorities like the government or the U.S. Federal Reserve System \added{as reliable sources}~(JP1, JP5, JR3). Regarding first-hand accounts, JP5 highlighted the importance of news organizations having their own photographers. JL1 argued that \final{when} anonymous sources are used, the information needs to be confirmed independently. JL5, who worked as a professional fact-checker, said that she always \final{got} experts on the phone to verify the information in a story and trace the provenance of claims. Considering the breadth of \final{one source's} reporting, JP2 wanted to know whether \final{the} website actually had reporters in all the countries they were covering or whether they pulled their information from other outlets. JP5 warned that few organizations can afford global reporting with people on the ground and experts on all topics.

\subsection{Reputation}

\final{\textit{End Users.}} 
\final{Eleven} of the 23 \final{end users}~(48\%) took the reputation of a source into account. They also commented on whether they already knew a source~(CDU3, SPD6), \final{and} relied on their experience doing research for their work as members of parliament~(SPD2, SPD4). CDU3 said he is ``extremely skeptical'' about sources he does not know. \final{End users} also referred to the reputation of the sources and took their personal experience with a source into account. SPD2 argued that the more abstruse a story is and the less familiar she is with a source, the less credible she considers the source. SPD6 \final{mentioned witnessing} ``an extensive discussion about [one of the sources] in the social left,'' \final{using} this to judge \final{the} sources. GN1 knew one of the sources because they wrote a ``stupid'' story about him \final{that resulted} in a ``shitstorm'' because other far-right websites picked up the story.

Many \final{end users~(CDU5, GN2, GN4, SPD3)} used search engines like Google to research news sources, \final{while} GN2 and GN4 used Wikipedia to inform themselves about a website's reputation. More generally, GN1 used other websites and references that he trusts and perceives as neutral and objective, GN4 said he would search \final{Twitter} to understand what people he knows and respects write about a source, \final{and} SPD2 argued that she would ask many people, including colleagues, experts, journalists, and the State Media Authority, before making her assessment. AFD2 and AFD3 \final{said they} would use their personal networks.

\final{\textit{Experts.}} 
\label{sec:reputation_experts}
\final{Eleven} of the 20 \final{experts}~(55\%) \final{considered} the reputation of a website and its authors. JC2 argued that ``somebody or something has to tell you that it is a credible website.'' JL3 considered the reputation of the sources used by a website. \final{For JR1, t}he fact that ad networks refused to work with one of the sources was ``setting off alarms.'' For the \added{economic} context, JC2 argued that if the writer \added{had} worked as a financial analyst for a large bank, she would have trusted him. 

\final{Experts also considered what they} had heard about a source~(JP4, JL1, JL4). If they knew the author, they took \final{into account} their opinion of \final{that} author and what they knew about the \final{author's} reporting and political bent~(JP4, JL4, JL6, JR2). \final{They also looked at} whether an author \final{was} well-established and knowledgeable~(JP1). \final{One author was described as} unreliable and politically motivated \final{by JR2, and as} ``intellectually dishonest'' by JP4. Despite sharing a political alignment, JR2 said he ``would never consider" the \final{author} a reliable source. JP3 knew one \final{website author, having previously} interviewed him. \added{Based on \final{this} professional interaction, he} felt ``very uncomfortable'' about using him in \final{a} story \final{he wrote in the past}. JR2 knew that information from one source \final{had been} overstated or falsely stated in the past, \final{and} found this problematic even though the website matched his understanding of the current economic market at the time of the \added{study}. 

Affiliation with reputable sources was seen \final{by experts} as an indicator of reliability~(JP1, JC2). \final{They} also considered what others, especially peers, thought about a source~(JL4, JL5, JR1, JR3). JL3 referred to the kinds of people \final{who} share an article as \final{an} indicator of reliability, \final{and JR1 noted} the perceived reliability of the peers and the sources \final{in which} they published. \final{Also considered was} how many followers \final{an author} had~(JR3), whether \final{experts} and experts in their personal social network followed, quoted, or shared content by the author~(JR3), whether a site was banned by Facebook~(JP3), and whether unreliable websites were presented in the related search tab on Google Search~(JL5).

The \final{experts} mentioned several ways of learning more about the reputation of a website~(JP1, JL1, JL4, JL5), e.g., using the name of the website and the term ``bias'' as keywords on Google Search~(JP1) and consulting Wikipedia~(JP1, JP3, JL5). The\final{y} also took into account what \final{was published about a source by} reliable or centrist mainstream sources like the New York Times~(JP1, JP3, JL1, JL5) and \final{by} institutions like the Southern Poverty Law Center~(JP3, JL5).

\subsection{Design of Website}

\final{\textit{End Users.}}  
\final{Twelve} of the 23 \final{end users}~(52\%) used \final{a website's} design and structure as criteri\final{a for evaluating its reliability, noting whether a site was cluttered and well-structured~(CDU4, CDU6, SPD4) and if the site and design ``looked reputable''~(CDU3, CDU6)}. SPD4 explained that, while fringe websites used to look like ``bad PowerPoint presentations,'' \final{today} many look more professional. \final{End users} \final{noted a site looking} ``wild'' or ``turbulent''~(CDU6, FDP1) \final{or resembling} a blog~(LT1, SPD4). Other design-related criteria include\final{d} whether a source was ``colorful''~(CDU6, FDP1), \final{the} fonts used~(SPD4), \final{consistency} (FDP1), the number of buttons (FDP1), and the overall impression \final{given by a site} (FDP1). CDU3 said that he ``immediately saw'' whether the first two sources were reliable. FDP1 and SPD2 considered \final{the similarity of} the design to \final{that of} the largest tabloid paper in Germany, \final{and} SPD3 \final{noted} how similar the design \final{was} to \final{that of} other unreliable sources.

\final{\textit{Experts.}} 
Far fewer professional \final{experts} commented on \final{a website's} design, \final{with only} 6 of the 20 ~(30\%) \final{using} design as a criterion. \final{Those who did noted} layout~(JP5, JP7, JC1, JC3), formatting~(JR1), fonts~(JC1), and relative size of images~(JC3) as indicators. JC1 also considered \final{similarity of the design to that of} reliable websites like the New York Times. \final{JC1 noted whether a website looked ``messy,'' ``clean,'' or ``serious''~(JC1), and JP3 criticized one website} for looking like a personal blog.

\subsection{Differences Between Experts and Non-Experts (RQ3)}

\final{An exploration of the differences} between experts and non-experts is important because it 1\final{)} helps us identify criteria \final{used by} people with experience \final{in} the production of news, 2\final{)} enables us to identify criteria within end users' abilities that are well-founded, and 3\final{)} allows us to debunk common \final{choices of} end users.

\added{Figure~\ref{fig:all_codes_small} indicates the proportion of \final{end users and experts} \final{who} used a criterion to \final{determine} whether a purported news website \final{was} reliable. Overall, we found strong similarities between the two groups, even though they reviewed different news sources in different languages and came from different countries and political backgrounds. We identif\final{ied} four criteria with little difference between \final{end users} and \final{experts:} political alignment (6\% difference), ownership (7\%), content (7\%), and reputation (7\%). Note that \final{content and political alignment are} the two most frequently used criteria\final{; we} see this as corroboration of the ecological validity of the criteria, \final{as these} criteria are frequently invoked by experienced news producers and \final{are} within end users' abilities.}

\final{Other} criteria \final{we}re considered helpful by one group but not  the other. The \final{top five} criteria for \final{end users} were advertisements, content, writing style, political alignment, and ownership, \final{while the top five for experts were} authors, content, political alignment, self-description, and professional standards. \final{For} two criteria the difference between the two groups \final{wa}s greater than 50\%. The criterion with the \added{most considerable} difference \final{was a}dvertising (67\%). While almost 20 of the 23 \final{end users} (87\%) took advertising into account, only 4 of the 20 \final{experts} (20\%) did so. \final{For the author criterion, with a 51\% difference, o}nly 9 of the 23 \final{end users} (39\%) referred to the authors, \final{compared to} 18 of the 20 experts (90\%). We believe that a likely explanation for this difference is that \final{experts as journalists} have experience being author\final{s, with a} better understanding of author\final{ship} and \final{the} influence \final{held by the} author of a news piece. \final{End users} who have never worked as journalists may be less familiar with how news websites are produced \final{and} may not fully realize this impact.

\section{Discussion}
\label{sec:discussion}

In this paper, we show that the overwhelming majority of \added{end users and experts} can identify unreliable websites~(RQ1), i.e., the ratings they \added{provide align with} what we and Gruppi et al.~\cite{gruppi2020nelagt2019} perceive as unreliable websites. This \added{agreement} is noteworthy, since \final{end users and experts had different skill levels and were from different countries with different media landscapes}. We also learned that the unreliable websites we used in this study were rarely visited by \added{participants}. \added{W}e identified eleven reliability criteria to assess whether a website is reliable~(RQ2) and found differences in the applicability of the criteria between politically diverse \final{end users} and \final{experts}~(RQ3).

\final{We acknowledge concerns voiced by scholars such as danah boyd that media literacy lessons may ``backfire'' and that even well-executed lessons on critical thinking can lead to ``ask[ing] people to doubt what they see,'' which can make them doubt both unreliable and reliable sources~\cite{boyd_2018_a}. Informed by these concerns, we focused on reliability criteria beyond simple heuristics that can be easily manipulated. Like boyd, we do not view media literacy as a panacea. Criteria such as the kind of content displayed on a website could be used to teach users what they should pay close attention to when evaluating the reliability of news websites.}

\final{The criteria presented in this paper can support the type of critical thinking and evaluation that Wineburg and McGrew refer to as lateral reading~\cite{wineburg2017lateral}. Our work corroborates and extends their lateral reading technique of scanning a website and consulting other websites~\cite{wineburg2017lateral} whose effectiveness is empirically shown~\cite{wineburg2022lateral}. Criteria such as political alignment and authorship can help users assess a source's political alignment and research an author's background.}

Table~\ref{tab:criteria_comparison} indicates which criteria are partially covered by prior work. In this discussion, we compare \final{our} criteria to \final{those seen in} prior work, introduce an analytical distinction \added{between manipulable and less manipulable criteria}, \final{and} describe how the \added{less manipulable criteria} can be augmented through technology.

\subsection{Comparison to Prior Work}
\label{sec:discussion_comparison_prior_work}

\begin{table}
\small
  \caption{\added{The eleven reliability criteria for news websites that we observed in practice based on contextual inquiry, think aloud study, and semi-structured interviews. We compare this to prior work based on surveys, discussions with stakeholders who are not the end users, or an unknown empirical basis.}}\label{tab:criteria_comparison}
  \Description{The table is a matrix that visualizes which criteria are covered by prior work. This is a visualization of the description in the text.}
{\color{black}\begin{tabular}{l|c|cc|cccc}
\toprule
Reliability Criteria & Survey & \multicolumn{2}{c|}{Expert Discussion} & \multicolumn{4}{c}{Not Explained} \\
based on: & \cite{10.1145/3415164} & \cite{zhang2018structured} & \cite{W3C} & \cite{bradshaw2020sourcing} & \cite{guess2020digital} & \cite{newsguard_2022} & \cite{TrustProject} \\
\midrule
\arrayrulecolor{lightgray}
01. Content & \checkmark & \checkmark &  &  & \checkmark & \checkmark & \\ \hline
02. Political Alignment &  &  &  &  &  &  & \\ \hline
03. Writing Style & \checkmark & \checkmark &  & \checkmark &  & \checkmark & \\ \hline
04. Authors &  &  &  &  &  & \checkmark & \checkmark \\ \hline
05. Self-Description &  &  &  &  &  &  & \checkmark  \\ \hline
06. Professional Standards & \checkmark &  &  & \checkmark &  & \checkmark & \checkmark \\ \hline
07. Advertisements &  & \checkmark &  &  &  & \checkmark &  \\ \hline
08. Ownership &  &  &  &  &  & \checkmark & \\ \hline
09. Sources & \checkmark & \checkmark &  &  & \checkmark & \checkmark & \checkmark \\ \hline
10. Reputation & \checkmark &  &  &  &  &  & \\ \hline
11. Design of Website & \checkmark &  &  & \checkmark & \checkmark &  & \checkmark \\
\arrayrulecolor{black}
\midrule
\midrule
Other Criteria & \checkmark & \checkmark  & \checkmark & \checkmark & \checkmark & \checkmark & \checkmark \\
\bottomrule
\end{tabular}}
\end{table}

\added{An important motivation for this work is to provide an empirical basis for reliability criteria. Table~\ref{tab:criteria_comparison} shows which \final{of the} criteria we observed are also mentioned in prior work. Unlike prior work, which based these criteria on surveys~\cite{10.1145/3415164} \final{or} discussions~\cite{zhang2018structured,W3C}, or which do not explain how the criteria were determined~\cite{bradshaw2020sourcing,guess2020digital,newsguard_2022,TrustProject}, our criteria are based on a user study \final{employing} think aloud methods and semi-structured interviews. The\final{se} criteria may be closer to what people use in practice. Compared to prior work, we recruited a larger and more diverse sample, \final{including} experts and non-experts and controll\final{ing} for \final{participants'} political alignment.}

\added{Table~\ref{tab:criteria_comparison} shows that some of the criteria we identified in practice are frequently covered in prior work, while others are rarely included. Frequently included criteria are \final{content, writing style, professional standards, sources, and website design}. Criteria covered by \final{only} one or two other lists include \final{authors, self-description, advertisements, ownership, and reputation}. We are the first to describe \final{political alignment}, which is surprising since it is one of the most frequently used criteria in our investigation. \final{The largest overlap is between our criteria and those of} NewsGuard~\cite{newsguard_2022}, covering \final{7 of the 11} criteria we identified. Six of \final{the 11 were} also covered by Bhuiyan et al. Our empirical study showed that criteria like \final{writing style and website design}, which are frequently included in criteria catalogs, are more frequently used by \final{end users than by experts}.}

\final{Another noteworthy finding is how few of our criteria are covered by those proposed by the Credible Web Community Group~\cite{W3C}. Their criteria focus on awards, correction policies, and how old a website is. In our study, neither end users nor experts used these criteria during their review of different websites.}

\added{Our comparison of \final{end users and experts} also showed that some criteria, like \final{authors, self-description, and professional standards}, are \final{used} frequently, especially by experts, but \final{are} rarely found in other compilations of reliability criteria. There is also no other work that covers all criteria.}

\subsection{\added{Criteria Where Manipulation Is \final{Less Difficult}}}

\added{The reliability criteria presented as results in Section~\ref{sec:results} are based on extensive empirical work on how purported news websites are evaluated in practice. \final{Here} we reflect on the generalizability of the\final{se} criteria and discuss how resilient the\final{y} might be against malicious actors. This is particularly important because adversaries will likely try to use the criteria for malicious intent. Before integrating reliability criteria into online platforms and other socio-technical systems aimed at supporting users, it is, therefore, \final{crucial} to critically examine \final{these criteria}. For this discussion, we distinguish between manipulable and less manipulable criteria. \final{While we} are aware that there may be exceptions for the less manipulable criteria, \final{b}ased on our experience from \final{43 semi-structured interviews with 23 end users from Germany and 20 experts from the United States,} we think that our classification can help guide efforts to support users.} \final{We recommend that the criteria for which manipulation is less difficult should not be taught to laypeople and should not be integrated into socio-technical tools, at least not without additional precautions to overcome manipulation risks.}

\added{The theoretical embedding of our classification is signaling theory~\cite{zahavi1975mate,spence1978job}, which is aimed at explaining why certain signals are reliable or not. Following \final{Donath}~\cite{10.1111/j.1083-6101.2007.00394.x}, signaling theory distinguishes assessment signals that are ``inherently reliable'' and conventional signals, where ``the link between signal and quality is arbitrary, a matter of social convention.'' An example of an assessment signal is the relationship between observing somebody lifting a heavy object and physical strength. An example of a conventional signal is self-descriptions in online social networks, where users can easily pretend to be older or younger.}

\added{Informed by the theoretical framework of signaling theory, we consider less manipulable criteria as assessment signals and more manipulable criteria as conventional signals. \final{Like Donath~\cite{donath2007signals}, we recognize that} very few signals are impossible to manipulate, given sufficient motivation. Nevertheless, introducing a distinction between manipulable criteria, which are comparatively easily manipulated, and less manipulable criteria, which are less difficult to manipulate, can help practitioners and others.}

Following an approach from cybersecurity~\cite{peikari2004security}, we asked ourselves ``What is the worst an attacker can do to you?'' and brainstormed how somebody operating an unreliable website might apply the criteria to appear more reliable. In this section, we present the criteria for which we identified ways in which the criteria can be manipulated.

The \final{w}riting \final{s}tyle of a website \final{was used by} both \final{end users and experts} to determine \final{a source's} reliability. \final{While} research indicates that writing style does affect \final{the} success \final{of} some misinformation and whether content has a strong chance of going viral~\cite{10.1145/3274351,potthast-etal-2018-stylometric}, \final{wording and writing style --- e.g., avoiding certain keywords and changing headlines --- do not affect the veracity of the content}. Therefore, writing style does not help people confidently assess whether a website is reliable.

Other criteria that can be easily manipulated include \final{a website's self-description and information about its ownership}. \added{Following signaling theory, the\final{se are conventional signals in which the link between reality and} what is presented about self-image or ownership is a mere social convention~\cite{10.1111/j.1083-6101.2007.00394.x}. \final{While end users and experts} used both criteria, \final{these are criteria that} can be easily \final{manipulated}. In our \added{study}, the claims made in \final{a website's} self-description and \final{information on} alleged ownership were \final{also} rarely verified by \added{participants}}. 

The \final{a}dvertisements \final{shown on a website were} another criterion \final{used,} especially \final{by} \final{end users}. \final{T}here \added{could}, \final{however,} be many unreliable websites that do not feature problematic advertisements, especially if those who run site\final{s} do not do \final{so} for personal gain, e.g., to push a political agenda. \final{While} forging this \added{criterion} would come at a cost for unreliable websites that are for-profit, it is too easily manipulated. \final{Website d}esign is \final{also} \added{easily manipulated}. \final{With} the availability of website builders and templates, it is comparatively easy to set up professional-looking websites, even for people without experience with HTML and CSS. \final{This criterion should therefore not be included in guidelines}.

\final{As illustrated by Figure~\ref{fig:all_codes_small}, the criteria for which manipulation is less difficult were those predominantly used by end users, including writing style, advertisements, and website design. One explanation for this could be users' limited understanding of news production processes. In their lived experience as end users, they may have experienced that reliable news is commonly presented in a professional writing style, and may also assume that advertisements and website design are difficult to manipulate, though this has become very easy.}

\subsection{\added{Criteria Where Manipulation Is \final{Difficult}}}

Based on our brainstorming of ways to deceive users, we \added{identify criteria \final{for which} manipulation is \final{difficult}, and invite others to join us in validating this empirically}. The first \added{such} criterion is the content covered on a website. As our empirical material showed, if a website spreads right-wing conspiracy theories or antisemitic content, this is hard evidence that a website is unreliable. Another aspect of the \final{c}ontent criterion is whether only one side of an issue is represented. \added{If an unreliable website tries to optimize for this criterion, e.g., by providing a balanced account that considers all sides of a discussion, this would improve the website and decrease its unreliability.}

Another \added{less manipulable} criterion is \final{a source's political alignment}. Akin to  content, \final{the featuring of} only certain political actors or ideas by a \added{news} website is hard evidence that \final{the} site is unreliable. A particularly useful and especially \added{hard-to-forge} aspect of political alignment is who is sharing links to a website on social media platforms like \final{X} \added{ or Mastodon, or on any other websites where people post text and links}. Since this is beyond the control of a website, it is \final{difficult} to manipulate. 

Which \final{a}uthors write for a source is another \added{criterion that is difficult to manipulate}. With available search engines, it is comparatively easy to research whether authors exist \final{and} have received proper training, and whether the content and opinions shown on a website are consistent with the author's prior work. 

Another important \added{criterion that is difficult to manipulate} is whether a website adheres to \final{p}rofessional \final{s}tandards. \final{This can be assessed by seeing if} a website clearly distinguishes between opinions and facts and \final{presents} different sides of a story. As our investigation showed, this can be done based on a topic that a user is already familiar with. 

\final{The s}ources \final{named} by a website can be easily checked, \final{allowing} the information provided --- \final{especially statistics --- to} be easily verified. \final{And the} sources used direct\final{ly} connect to the \final{criterion of reputation, which is also difficult} to manipulate. \final{A} reliable third party like Wikipedia or Snopes vot\final{ing} to classify a source as unreliable \final{can be considered hard evidence of that unreliability, as}  it would be \final{difficult} to manipulate such third-party assessments.

\final{Figure~\ref{fig:all_codes_small} indicates that the most frequently used criteria, content and political alignment, are used equally frequently by both end users and experts. There are three criteria that are difficult to manipulate and more frequently used by experts: authors, professional standards, and sources. One possible explanation as to why these three are used more frequently is the lived experienced of experts.}

\subsection{Implementing the Reliability Criteria for News Websites}

\begin{table}
\small
  \caption{\added{The eleven reliability criteria for news websites observed in practice via contextual inquiry, a think aloud study, and semi-structured interviews. We highlight which criteria we consider challenging to manipulate based on our analysis.}}\label{tab:criteria_manipulation}
  \Description{The table is a matrix that visualizes which criteria are covered by prior work. This is a visualization of the description in the text.}
{\color{black}\begin{tabular}{l|c}
\toprule
Reliability Criteria &  Manipulation is difficult \\
\midrule
\arrayrulecolor{lightgray}
01. Content & \checkmark \\ \hline
02. Political Alignment & \checkmark \\ \hline
03. Writing Style &  \\ \hline
04. Authors & \checkmark \\ \hline
05. Self-Description &   \\ \hline
06. Professional Standards & \checkmark \\ \hline
07. Advertisements &  \\ \hline
08. Ownership & \checkmark \\ \hline
09. Sources & \checkmark \\ \hline
10. Reputation & \checkmark \\ \hline
11. Design of Website &  \\
\arrayrulecolor{black}
\bottomrule
\end{tabular}}
\end{table}

Table~\ref{tab:criteria_manipulation} shows that \final{content, political alignment, authors, professional standards, sources, and reputation} are the six criteria that\added{, based on signaling theory and \final{on} our theoretical insights on how to manipulate \final{these criteria}, we consider to be most helpful in that they are a) difficult to manipulate in practice, b) used by experts, and c) applicable by \final{end users}. We therefore think that these criteria should be evaluated further. After successful \added{empirical} evaluations, they could be taught widely in media literacy and media education contexts.}

\added{The following section describe\final{s} how HCI can help users assess these criteria.} Our goal is to augment users' ability to recognize unreliable online information. \final{T}he technical proposals presented in this section, \final{however,} are only part of a larger socio-technical assemblage, \final{and depend} on users who are aware of the complexities of misinformation. For these users, they could become a powerful support.

\textit{Content.}
Assessing content and identifying \added{how and why} controversial topics are presented is challenging, but not without precedent. Media scholars like Puschmann et al.~\cite{doi:10.1177/13548565221109440} have employed word lists to successfully identify popular conspiracy tropes and topics like antisemitism, anti-elitism, anti-immigration/\final{I}slamophobia, and anti-gender/anti-feminism. Such word lists, \final{along with} more data-driven approaches like topic modeling via Latent Dirichlet Allocation (LDA), could be used to build HCI tools that enable users to assess whether a source is covering content that is considered problematic. \final{T}he word lists could also serve as a starting point for more in-depth qualitative investigations by fact-checkers.

\textit{Political Alignment.}
People are politically affiliated with a source if they \added{regularly} share it. \added{Data mining of social media platforms could be used to locate different sources in the political spectrum}. \added{\final{P}oliticians and what they share on \final{X}, Mastodon, or any other website could be tracked, \final{along with the} links shared by users and \final{information on} who users follow. \final{These} data mining approaches} would allow people to identify sources with multipartisan approval consumed across different political parties. To reliably detect extreme content, the social media profiles of people with a reputation for sharing misinformation~\final{--- e.g., Alex Jones in the U.S. or Attila Hildmann in Germany~---} could be included. Such HCI tools would make it comparatively easy for people to assess whether a website has multipartisan approval or whether \final{it} is only popular with a \added{specific} political group.

\textit{Authors.}
The author criterion could be supported by specialized search engines that make it easier to understand the authors of a website and their reputation. Thanks to social media, it has become harder to invent authors. Follower networks on platforms like \added{LinkedIn,} \final{X}, \added{Mastodon,} Instagram, or TikTok make it easier to investigate authors' connections and political affiliations.

\textit{Professional Standards.}
\final{S}pecialized search engines could help users investigate whether a source has been criticized in the past for not adhering to professional standards. Software tools could also retrieve sections titled ``Criticism'' or ``Controversy'' in the Wikipedia article \final{about} a website.

\textit{Sources.} The assessment of whether a news website \added{provides} sources for claims is another criterion that could be supported through technology. HCI researchers could build tools that leverage techniques like named entity recognition to identify the people and institutions cited in an article. \added{These names could then be used to retrieve the authors' and institutions' social media profiles \final{which}, along with information from Wikipedia articles and similar sources, could be used to ensure that \final{an} article's statements are consistent with what \final{the} authors and institutions have said in the past.} HCI tools could also apply more sophisticated approaches like argumentation mining to identify the key arguments in a text and  assess the\final{ir} quality ~\cite{wachsmuth-etal-2017-computational,skitalinskaya-etal-2021-learning}. If the key arguments are not accompanied by actors that support the\final{m}, this can be highlighted as problematic. Such tools could be valuable for fact-checkers, \final{and} less experienced users could be trained \final{in media literacy classes} to use the tools.

\textit{Reputation.}
A data-driven approach could make it easier to assess the reputation of a website using existing resources. This approach \final{could be} similar to \final{that of} prior work that helps people fact-check claims~\cite{10.1145/3491102.3517717}. \final{Platforms like Wikipedia are interesting in this context because} the community \final{uses a process of deliberation to} decide which sources to ban. \added{Using the strategies of JP1 described} in Section~\ref{sec:reputation_experts}, users could also enter the name of a website and keywords like ``bias'' into a search engine to see whether other reliable sources \final{have} accused a website of being biased. More generally, GN4's strategy of searching \final{social media websites to understand what kind of users post links to a news websites could be supported through HCI tools}.

\subsection{Political Alignment}

As \final{we have shown}, political alignment has not been covered in prior work, though it was widely used in our study and \final{it is} comparatively \final{difficult} to manipulate for attackers who create misinformation. \final{This lack of recognition} is surprising, \final{as} it directly relates to why people create misinformation~\cite{wardle_2018}. Political alignment connects to political influence, propaganda, and partisanship. \final{W}e found that political alignment is \final{frequently taken into account by} both \final{end users} and experts when rating the reliability of news websites or other information sources. One explanation why political alignment is not included by commercial businesses like Facebook~\cite{doi:10.1073/pnas.1920498117} and NewsGuard Criteria~\cite{newsguard_2022} is that it is in their business interest to appear neutral. Platforms like Facebook and \final{X} have repeatedly been criticized as being biased \added{in the past}. Asking users to take \final{a news website's} political alignment into account could, therefore, make platforms vulnerable to additional scrutiny and controversy. However, as Austrian-American communication theorist Paul Watzlawick famously argued, ``One cannot not communicate.'' \final{I}naction on the issue of political alignment, \final{therefore,} is also an action. By ignoring the influence of political alignment, Facebook, NewsGuard, and others could enable certain actors to push propaganda and gain political influence. \final{We do acknowledge, however,} that labeling political alignment is a challenging task \added{related} to the broader debate around democratic platforms and governance.

\subsection{Limitations}
\label{sec:limitations}

\added{The goal of \final{our} user study was to elicit criteria that can be used to determine whether a \final{news} website is reliable. The framing of \final{our} study represents a special setting in which users take time to pause and reflect on \final{a website's} reliability. As discussed, prior work indicates that misinformation is sometimes shared because people do not pay attention~\cite{pennycook2019lazy,pennycook2021shifting}. Further work is needed to understand how representative \final{our} study setting is for users who encounter unreliable news websites in their everyday life. Regardless of the outcome of such studies, the news reliability criteria presented in this paper can be used to sensitize users to pay attention to online websites and train them \final{to evaluate reliability}.}

\added{Participants' views on news reliability are shaped by their countries, cultures, and the political systems \final{in which they live}. It is also important to note that notions of ``left'' and ``right'' are always relative to \final{a specific country}.}

\added{Due to the methodological complexity and time limitations when consulting many \final{end users and experts}, we could only examine three news websites. \final{We thus} focused on unreliable websites \final{so as} to maximize our findings' impact on helping people identify misinformation. We encourage future work that applies the methodology to reliable websites.}

\added{\final{For end users}, we recruited elected politicians of a state parliament \final{so we could} include the expertise and perspectives \final{of people from} all political alignments. \final{While} these \final{end users} \final{we}re elected in free and open elections and, in theory, represent\final{ative} of the voting body, we found that \final{they} were highly educated. This level of education corresponds to prior work \final{showing} that highly educated people are overrepresented in legislative bodies around the world~\cite{erikson2019does}. \final{Still}, the \final{end users} in our investigation \final{were} laypeople regarding how news is produced. Our results are, therefore, helpful \final{in} understand\final{ing} \final{end users} without expertise in news production. That said, future work is needed to better understand the influence of education in this regard.}\\

\section{Conclusions}

A politically informed citizenry is the backbone of a well-functioning democracy~\cite{DelliCarpini1997}. We believe that the \added{news reliability} criteria described in this paper are an important stepping stone toward a more informed citizenry. \added{The\final{se} criteria are empirically grounded in that \final{end users and experts} apply them in practice. Such criteria have three benefits: 1) solutions based on these criteria are scalable since the assessment only has to be done for each news source, not each article, 2) the assessment of a news source can be done independently of individual claims and news articles, thus minimizing the risk of potential backfire effects, and 3) many criteria may also be applicable to individual news articles and other kinds of content \final{such as} videos. Another important benefit of news reliability criteria is that} they empower users to make their own decisions. History has shown, numerous times, that censorship and media control are central instruments of tyranny and dictatorship. Therefore, the most \added{critical} design goal of news reliability criteria is freedom of speech. None of the criteria in this paper tell people what to think or do, \final{and all of them} can help people make better and more informed decisions about the content they encounter online. \added{We make concrete technical proposals that can empower people to determine whether a news website is reliable.} With \added{our} criteria, we believe that platforms like YouTube, TikTok, Facebook, \final{X}, \added{Mastodon}, NewsGuard, and others can improve their services. Unlike other approaches towards misinformation connected to the risk of censorship, the criteria presented in this paper provide people with strategies for distinguishing reliable from unreliable sources in a way independent of political ideology. We hope that social media platforms and providers of reliability criteria make the\final{se} criteria available to their users.

\begin{acks}
This research has been supported by the German Academic Exchange Service~(DAAD), the German Research Foundation, Grant No.~374666841 (SFB 1342), and the NSF under Grant No.~2107391. We thank all participants for their time and their insights. We also thank the reviewers for their exceptionally helpful and constructive feedback.
\end{acks}

\bibliographystyle{ACM-Reference-Format}
\bibliography{references}

\appendix

\section{Appendix}

\begin{table}[h]
\small
  \caption{The politically diverse \final{end users} \added{from Germany that participated}. The table shows their party, gender, whether they have personally used any of the websites they rated, and how they rated them. Rating scale: very unreliable~($--$), unreliable~($-$), neither reliable or unreliable~($\medcirc$), reliable~($+$), very reliable~($++$).}~\label{tab:participants_and_ratings_de}
  \Description{The table provides all individual ratings of the 23 \final{end users}. Except for the small number of cases described in the text, the large majority of \final{end users} never used any of the three websites. \final{End users} rarely rated websites as reliable or very reliable.}
\begin{tabular}{ll|lll|ccc}
\toprule
    \multicolumn{2}{c|}{Participant}   & \multicolumn{3}{c|}{Personal Usage}      & \multicolumn{3}{c}{Reliability Ratings} \\
    ID    &  Gender  & DE1 & DE2 & DE3 & DE1     & DE2    & DE3    \\
\midrule
AFD1   & m & \textbf{Sometimes}            & Never                     & \textit{Rarely}                     & $\textbf{+}$    &               &                  \\
AFD2   & m & \textbf{Frequently}                 & Never                     & \textit{Rarely}                     & $\medcirc$             & $\medcirc$                & $\medcirc$                   \\
AFD3   & m & Never                 & Never                     & Never                        & $\medcirc$             & $--$         & $\medcirc$                   \\
CDU1   & m & Never                 & Never                     & Never                        & $--$      & $--$         & $\medcirc$                   \\
CDU2   & f & \textbf{Sometimes}            & Never                     & Never                        &            & $\medcirc$                & $-$                 \\
CDU3   & m & Never                 & Never                     & Never                        & $\medcirc$             & $\medcirc$                & $\medcirc$                   \\
CDU4   & m & Never                 & Never                     & \textit{Rarely}                     & $\medcirc$             & $\textbf{+}$       & $-$                 \\
CDU5   & m & Never                 & Never                     & Never                        & $-$           & $\medcirc$                & $-$                \\
CDU6   & f & Never                 & Never                     & Never                        & $\medcirc$             & $-$              & $-$                 \\
CDU7   & f & Never                 & Never                     & Never                        & $-$                    & $--$             & $--$                \\
FDP1   & m & Never                 & Never                     & Never                        & $-$           & $\medcirc$                & $-$                 \\
GN1 & m & Never                 & Never                     & Never                        & $--$      & $--$         & $--$            \\
GN2 & m & Never                 & Never                     & Never                        & $-$           & $-$              & $-$                 \\
GN3 & f & Never                 & Never                     & Never                        & $\medcirc$             & $\textbf{+}$       & $-$                 \\
GN4 & m & Never                 & Never                     & Never                        & $--$      & $--$         & $--$            \\
LT1  & f & Never                 & \textit{Rarely}                  & Never                        & $-$           & $-$              & $-$                 \\
LT2  & f & Never                 & Never                     & Never                        & $-$           & $--$         & $--$            \\
SPD1   & m & Never                 & \textbf{Frequently}                     & Never                        & $-$           & $\textbf{+}$       & $-$                 \\
SPD2   & f & \textit{Rarely}              & Never                     & Never                        &            & $-$              & $-$                 \\
SPD3   & m & \textit{Rarely}              & \textit{Rarely}                  & Never                        & $\medcirc$             & $-$              & $-$                 \\
SPD4   & m & Never                 & Never                     & Never                        & $--$      & $--$         & $--$            \\
SPD5   & f & Never                 & Never                     & Never                        & $-$           & $--$         & $-$                 \\
SPD6   & m & Never                 & \textit{Rarely}                  & Never                        & $\medcirc$             & $\medcirc$                & $--$           \\
\bottomrule
\end{tabular}
\end{table}

\begin{table}[h]
\small
  \caption{\added{The professional journalists from the U.S. that participated \final{as experts}}. The table shows their gender, whether they have personally used any of the websites they rated, and how they rated them. \added{The ID indicates what kind of newspaper \final{employs which expert}. For \final{experts} from the Top 10 most popular US-based newspapers, the participant ID begins with JP. For \final{experts} recruited from newspapers with a political alignment, JL stands for left, JC for center, and JR for right.} Rating scale: very unreliable~($--$), unreliable~($-$), neither reliable or unreliable~($\medcirc$), reliable~($+$), very reliable~($++$).}~\label{tab:participants_and_ratings_en}
  \Description{The table shows all individual ratings by the 20 \final{experts}. Except for the small number of cases described in the text, the large majority of \final{experts} never used any of the three websites. \final{Experts} also rarely rated websites as reliable or very reliable.}
\begin{tabular}{ll|lll|ccc}
\toprule
    \multicolumn{2}{c|}{Participant}   & \multicolumn{3}{c|}{Personal Usage}      & \multicolumn{3}{c}{Reliability Ratings} \\
    ID    &  Gender  & EN1 & EN2 & EN3 & EN1     & EN2    & EN3    \\
\midrule
JP1        & f               & Never           & Never        & Never        & $-$          & $-$             & $--$         \\
JP2        & f               & Never           & Never        & Never        & $--$         & $--$            & $\medcirc$   \\
JP3        & m               & Never           & Never        & Never        & $--$         & $-$             & $--$         \\
JP4        & m               & Never           & Never        & Never        & $--$         & $--$            & $-$          \\
JP5        & m               & Never           & Never        & Never        & \textbf{$+$} & $-$             & $--$         \\
JP6        & m               & Never           & Never        & Never        & $-$          & $--$            & $-$          \\
JP7        & m               & \textit{Rarely} & Never        & Never        & $\medcirc$   & $-$             & $-$          \\
\midrule
JL1        & m               & Never           & Never        & Never        & $--$         & $--$            & $--$         \\
JL2        & m               & \textit{Rarely} & Never        & Never        & $-$          & $--$            & $--$         \\
JL3        & f               & Never           & Never        & Never        & $--$         & $-$             & $\medcirc$   \\
JL4        & m               & Never           & Never        & Never        & $--$         & $--$            & $--$         \\
JL5        & f               & Never           & Never        & Never        & $-$          & $--$            & $--$         \\
JL6        & m               & Never           & Never        & Never        & $--$         & $--$            & $--$         \\
\midrule
JC1        & f               & Never           & Never        & Never        & $\medcirc$   & \textbf{$+$}    & $\medcirc$   \\
JC2        & f               & Never           & Never        & Never        & $-$          & $\medcirc$      & $-$          \\
JC3        & f               & Never           & Never        & Never        & $--$         & $--$            & $--$         \\
\midrule
JR1        & f               & Never           & Never        & Never        & $--$         & $--$            & $--$         \\
JR2        & m               & \textit{Rarely} & Never        & Never        & $\medcirc$   & $-$             & $--$         \\
JR3        & f               & Never           & Never        & Never        & $\medcirc$   & $--$            & $--$         \\
JR4        & m               & Never           & Never        & Never        & $\medcirc$   & $\medcirc$      &              \\
\bottomrule
\end{tabular}
\end{table}

\end{document}